 [2005/06/22 5.2/AAS markup document class]%
\documentclass[preprint2]{aastex}%

\makeindex

\makeatletter 

\let\o@verbatim\verbatim
\def\verbatim{%
  \ifhmode\unskip\par\fi
  \ifx\@currsize\normalsize
     \small
  \fi
  \o@verbatim
}

\renewcommand \verbatim@font {%
  \normalfont \ttfamily
  \catcode`\<=\active
  \catcode`\>=\active
}

\RequirePackage{shortvrb}
\MakeShortVerb{\|}

\begingroup
  \catcode`\<=\active
  \catcode`\>=\active
  \gdef<{\@ifnextchar<\@lt\@meta}
  \gdef>{\@ifnextchar>\@gt\@gtr@err}
  \gdef\@meta#1>{\m{#1}}
  \gdef\@lt<{\char`\<}
  \gdef\@gt>{\char`\>}
\endgroup
\def\@gtr@err{%
   \ClassError{ltxguide}{%
      Isolated \protect>%
   }{%
      In this document class, \protect<...\protect>
      is used to indicate a parameter.\MessageBreak
      I've just found a \protect> on its own.
      Perhaps you meant to type \protect>\protect>?
   }%
}
\def\verbatim@nolig@list{\do\`\do\,\do\'\do\-}

\newcommand{\m}[1]{\mbox{$\langle$\it #1\/$\rangle$}}

\def\cmd#1{\cs{\expandafter\cmd@to@cs\string#1}}
\def\cmd@to@cs#1#2{\char\number`#2\relax}
\DeclareRobustCommand\cs[1]{\texttt{\char`\\#1}}
\def\GetFileInfo#1{%
  \def\filename{#1}%
  \def\@tempb##1 ##2 ##3\relax##4\relax{%
    \def\filedate{##1}%
    \def\fileversion{##2}%
    \def\fileinfo{##3}}%
  \edef\@tempa{\csname ver@#1\endcsname}%
  \expandafter\@tempb\@tempa\relax? ? \relax\relax}

\def\mybf{}
\def\respbf{}
\def\newbf{}

\makeatother
\title{%
 Probing the Mysteries of the X-Ray Binary 4U 1210-64\\ 
 with ASM, PCA, MAXI, BAT and Suzaku
}%
\title{\today}

\author{Joel~B.~Coley\altaffilmark{1,2},
Robin~H.~D.~Corbet\altaffilmark{1,2},
Koji~Mukai\altaffilmark{1,2},
Katja~Pottschmidt\altaffilmark{1,3}}
\email{jcoley1@umbc.edu}
\altaffiltext{1}{University of Maryland Baltimore County, 1000 Hilltop Cir, Baltimore, MD 21250 USA}
\altaffiltext{2}{CRESST/Mail Code 662, X-ray Astrophysics Laboratory, NASA Goddard Space Flight Center, Greenbelt, MD 20771, USA}
\altaffiltext{3}{CRESST/Mail Code 661, X-ray Astroparticle Physics Laboratory, NASA Goddard Space Flight Center, Greenbelt, MD 20771, USA}

\expandafter\GetFileInfo\expandafter{\jobname.tex}%

\begin{document}

\begin{abstract}

4U 1210-64 has been postulated to be a High-Mass X-ray Binary powered by the Be mechanism. X-ray observations {\respbf with} \textsl{Suzaku}, the \textsl{ISS} Monitor of All-sky X-ray Image (MAXI) and the {\respbf \textsl{Rossi X-ray Timing Explorer}} Proportional Counter Array (PCA) and All Sky Monitor (ASM) provide detailed temporal and spectral {\respbf information} on this poorly understood source. Long term ASM and MAXI observations show distinct high and low states and the presence of a 6.7101$\pm$0.0005\,day modulation, {\respbf interpreted as the orbital period}. Folded light curves reveal a sharp dip, {\respbf interpreted as an eclipse. To determine the nature of the mass donor}, the predicted eclipse half-angle was calculated as a function of inclination angle for several stellar spectral types. The eclipse half-angle is not consistent with a mass donor of spectral type B5 V; {\respbf however, stars with spectral types B0 V or B0-5 III are possible. The best-fit spectral model consists of a power law with index $\Gamma$=1.85$^{+0.04}_{-0.05}$ and a high-energy cutoff at 5.5$\pm$0.2\,keV modified by an absorber that fully covers the source as well as partially covering absorption.} Emission lines from S XVI K$\alpha$, Fe K$\alpha$, Fe XXV K$\alpha$ and Fe XXVI K$\alpha$ were observed in the \textsl{Suzaku} spectra. Out-of-eclipse, Fe K$\alpha$ line flux was strongly correlated with unabsorbed continuum flux, {\respbf indicating} that the Fe I emission is the result of fluorescence of cold dense material near the compact object. The Fe I feature is not detected during eclipse, further supporting an origin close to the compact object.

\end{abstract}

\section{Introduction}

4U 1210-64 is an unusual X-ray Binary (XRB) first detected by the \textsl{Uhuru} satellite in 1978 \citep{1978ApJS...38..357F}.  \citet{1978ApJS...38..357F} found that 4U 1210-64 appears to be a variable source with a mean flux of 8.9$\times$10$^{-11}$\,erg cm$^{-2}$ s$^{-1}$ in the 2-6\,keV band.  Subsequent detections of 4U 1210-64 include the \textsl{Einstein} \citep{1992ApJS...80..257E} and \textsl{EXOSAT} \citep{1999A&AS..134..287R} slew surveys at fluxes 1.7$\times$10$^{-11}$\,erg cm$^{-2}$ s$^{-1}$ and $\sim$7$\times$10$^{-10}$\,erg cm$^{-2}$ s$^{-1}$ in the 0.16--3.5\,keV and 1--8\,keV bands, respectively; the wide-field cameras (WFC) on board \textsl{BeppoSAX} \citep{2007A&A...472..705V}, \textsl{INTEGRAL}/IBIS \citep{2003A&A...411L.131U} and the Burst Alert Telescope (BAT) on board the \textsl{Swift} observatory \citep{2010ApJS..186..378T, 2010A&A...510A..48C} at fluxes 2.57$\times$10$^{-10}$\,erg cm$^{-2}$ s$^{-1}$, 1.1$\times$10$^{-11}$\,erg cm$^{-2}$ s$^{-1}$ and $\sim$2$\times$10$^{-11}$\,erg cm$^{-2}$ s$^{-1}$ in the 2--10\,keV, 20--100\,keV and 14--195\,keV bands, respectively.

A soft X-ray counterpart at coordinates (J2000) RA=12:13:14.7, Dec=-64:52:31 with a position uncertainty of $\sim$4$\arcsec$ was revealed \citep{2007ATel.1253....1R} in observations conducted with the X-ray telescope \citep[XRT,][]{2005SSRv..120..165B} on board \textsl{Swift}.  \citet{2007ATel.1253....1R} modeled the X-ray spectrum of 4U 1210-64 using a continuum consisting of a power law with hard photon index and an emission feature at 6.7\,keV with an equivalent width (EQW) of $\sim$400 eV.  These observations resulted in a classification of 4U 1210-64 as an intermediate polar cataclysmic variable \citep[CV,][]{2007ATel.1253....1R}.

Observations using the \textsl{Swift}/XRT in late 2006 and early 2008 confirmed the presence of an emission feature at 6.7\,keV, thought to be Fe XXV \citep{2010ApJ...511..A48}.  The emission feature, observed to be prominent when the source is at intermediate flux levels, was not seen when the source entered periods of low or high flux \citep{2010ApJ...511..A48}. At intermediate flux levels, the EQW of the emission line was observed to be $\sim$1.6\,keV.  This very large EQW is suggestive of a blend of Fe lines.  While \citet{2007ATel.1253....1R} suggested that the spectral properties indicate that 4U 1210-64 is a CV, \citet{2010ApJ...511..A48} proposed that the stellar remnant in the system is a neutron star based on the very large EQW and variability of the emission feature.  Under the assumption of a power law continuum and a distance of $\sim$2.8\,kpc \citep{2009A&A...495..121M}, the high state X-ray luminosity of 4U 1210-64 was found to be 1.9$\times$10$^{35}$ erg s$^{-1}$ in the 2--10\,keV band \citep{2010ApJ...511..A48}.  This exceeds the typical luminosities observed in CVs by a factor of $\sim$1--2 orders of magnitude \citep{2006MNRAS.372..224B, 2008A&A...489.1121R, 2009A&A...496..121B}. A blackbody soft excess with a temperature of $\sim$1.5\,keV was found in 4U 1210-64, which implies that the accretor is more compact than a white dwarf \citep{2010ApJ...511..A48}.  \citet{2010ApJ...511..A48} concluded that the presence of the soft excess provides compelling evidence against a CV interpretation of the system.

The optical counterpart of 4U 1210-64 was observed {\mybf on MJD\,54529.3} using the 1.5\,m Cerro 
Tololo Interamerican Observatory (CTIO) in Chile \citep{2009A&A...495..121M}.  Based on its 
optical spectrum, \citet{2009A&A...495..121M} proposed that the spectral class of the mass donor 
is B5 V.  The features observed in the optical spectrum consist of Balmer series lines in 
absorption and emission of neutral helium, singly ionized helium and a blend of doubly ionized 
nitrogen and carbon.  The EQW of the H$\alpha$ line was observed to be approximately 2.5\,${\rm \AA}$ \citep{2009A&A...495..121M}.  This is less than half the expected value for a B5 V star, 
which suggests that emission features are present.  While an early-type mass donor suggests that 
4U 1210-64 is a High-Mass X-ray Binary, the presence of a B5 V main-sequence star in a HMXB would 
be unprecedented \citep{1998A&A...338..505N}.  The majority of Be X-ray binaries (BeXBs) host 
primaries with spectral properties that range from late O to early B type stars.  This will be 
discussed in further detail in Section 4.

A 6.7 day orbital period was discovered using data from the \textsl{Rossi X-ray Timing Explorer} All-Sky Monitor (ASM) \citep{2008ATel.1861....1C}.  In addition, \citet{2008ATel.1861....1C} report three main states of the system along with the possibility of an eclipse.  We investigate these results in more detail in Section 3.1 using additional data from ASM, the Monitor of All-Sky X-ray Image (MAXI) on board the \textsl{International Space Station (ISS)} and Burst Alert Telescope (BAT) instruments on board the \textsl{Swift} spacecraft.

This paper is structured in the following order:  \textsl{Suzaku}, MAXI, PCA and ASM observations are presented in Section 2; Section 3 focuses on the results of the X-ray campaign and Section 4 presents a discussion of the results.  The conclusions are outlined in Section 5.  If not stated otherwise, the uncertainties and limits presented in the paper are at the 1$\sigma$ confidence level.

\section{Observations and Data Analysis}

The observations outlined below consist of data collected during a two day \textsl{Suzaku} observation of 4U 1210-64 (2010 Dec. 23--25), pointed observations using the \textsl{RXTE} Proportional Counter Array (PCA) as well as long-term observations of the system using the following all-sky monitors: MAXI, ASM, and BAT.

\subsection{RXTE}

\subsubsection{ASM}

The ASM on board RXTE \citep{1996ApJ...469L..33L} consisted of three coded-aperture Scanning Shadow Cameras (SSCs), each containing a position-sensitive proportional counter with a field-of-view (FOV) of 6$\arcdeg{\times}$90$\arcdeg$ FWHM \citep{1997asxo.proc...29R}.  ASM scanned approximately 80$\%$ of the sky per spacecraft orbit ($\sim$90 minutes) with a 90\,s time resolution (``dwells") \citep{1997asxo.proc...29R, 1996ApJ...469L..33L}.  The FOV of the cameras is fixed on the sky for the duration of a ``dwell".  Sensitive to energies in the 1.5--12\,keV band, ASM observed 4U 1210-64 from MJD 50087 to 55924.

The light curves were retrieved from the ASM/\textsl{RXTE} database\footnote{http://xte.mit.edu/ASM\_lc.html} managed by MIT, which includes ``dwell-by-dwell" light curves and daily averaged light curves.  The light curves are divided into three energy bands: 1.5--3\,keV, 3--5\,keV and 5--12\,keV. Over the entire energy range (1.5--12\,keV), the Crab produces approximately 75.5\,counts s$^{-1}$ \citep{1997asxo.proc...29R, 1996ApJ...469L..33L}. ``Blank field" observations of regions at high Galactic latitudes indicate that a systematic uncertainty of $\sim$0.1\,counts\,s$^{-1}$ must be taken into account \citep{1997asxo.proc...29R, 1996ApJ...469L..33L}.  We used the ``dwell-by-dwell" light curves in our analysis (see Section 3.1).

\subsubsection{PCA}
\label{PCA Description}

Consisting of five proportional counter units (PCUs), the PCA was sensitive to X-rays in the energy band 2--60\,keV.  The total effective area of the instrument was $\sim$6500\,cm$^2$ with a FOV at FWHM of 1$\arcdeg$ \citep{1996SPIE.2808...59J}.  {\mybf At 6.0\,keV, the energy resolution FWHM in the PCA is $\sim$1\,keV \citep{2006ApJS..163..401J}.}

4U 1210-64 was observed 65 times between MJD\,54804.0--54842.0.  Individual observations typically lasted for 2--4\,ks but were as short as 1.4\,ks and as long as 45.6\,ks.  Dataset IDs were 93455-01 (23 observations) and 94409-01 (42 observations). The majority of the observations were performed during a single spacecraft orbit with no interruptions; longer ones spanned multiple orbits, with interruptions due to Earth occultations and/or SAA passages. Of the 5 PCUs, only PCU2 was used consistently for these observations, so we have opted to analyze only data taken with PCU2.  The Crab produces $\sim$2000--2400\,counts s$^{-1}$ in the PCU2 top layer\footnote{http://www.sternwarte.uni-erlangen.de/wilms/rxte/}.

Each detected event was recorded in different ways by the on-board experiment data system (EDS).  We analyzed Standard2 mode data, producing 129-channel spectra every 16\,seconds, for spectral analysis and for low-frequency timing analysis.  We generated the ``faint'' model background\footnote{http://heasarc.nasa.gov/docs/xte/recipes/pcabackest.html}, since the source mostly stayed below 40\,counts s$^{-1}$ PCU$^{-1}$. For high-frequency timing analysis, we generated light curves in 10\,ms bins using the GoodXenon event mode data, without background subtraction.  Spectral data including background subtraction were reduced and analyzed using the standard screening criteria \citep{2006ApJS..163..401J}.

\subsection{MAXI}
The MAXI instrument is an X-ray slit camera sensitive to energies 0.5--30\,keV \citep{2010SPIE.7732E..26M}.  MAXI consists of two types of slit cameras, the Gas Slit Camera (GSC) and the Solid-state Slit Camera (SSC), which observe the X-ray variability of over 1000 sources over every ISS orbit of approximately 92 minutes \citep{2010SPIE.7732E..26M}.  Of the two cameras, we make use of the GSC data that are routinely made available by the MAXI team.

Covering an energy range of 2--30\,keV, the GSC consists of twelve one-dimensional position sensitive proportional counters (PSPC), which make six camera units \citep{2010SPIE.7732E..26M}.  The overall FOV is a 160$\arcdeg{\times}$3$\arcdeg$ slit, which scans both the horizon and zenith directions \citep{2010SPIE.7732E..26M}.

We analyzed MAXI data obtained between MJD 55061.5--56394.5.  Light curves of energies 2--4\,keV, 4--10\,keV and 10--20\,keV were retrieved from the data available in the MAXI RIKEN database\footnote{http://maxi.riken.jp/top/}.  The 2--4\,keV and 4--10\,keV light curves were subsequently co-added for comparison with the light curves produced by the ASM.

\subsection{Swift}

The BAT on board the \textsl{Swift} spacecraft is a hard X-ray telescope operating in the 15--150\,keV energy band \citep{2005SSRv..120..143B}.  The detector is composed of CdZnTe where the detecting area and field of view (FOV) are 5240\,cm$^2$ and 1.4\,sr (half-coded), respectively \citep{2005SSRv..120..143B}.  The BAT provides an all-sky hard X-ray survey with a sensitivity of $\sim$2\,mCrab \citep{2005SSRv..120..143B}.  The Crab produces $\sim$0.045\,counts s$^{-1}$ over the 14--195\,keV energy band.

We analyzed BAT data obtained during the time period MJD 55152--56141.  Light curves were retrieved using \citet{2013ApJS..209...14K}'s extraction of the BAT data available on the NASA GSFC HEASARC website\footnote{http://heasarc.gsfc.nasa.gov/docs/swift/results/transients/}, which includes orbital and daily-averaged light curves.  We used the orbital light curves in the 15--50\,keV energy band in our analysis (see Section 3).

\subsection{Suzaku}

The \textsl{Suzaku} observation of 4U 1210-64 took place in 2010 December 23--25 (MJD 55553.16--55555.15) with an exposure time of $\sim$80\,ks (ObsID 405045010).  Data, collected using the X-ray Imaging Spectrometer (XIS) and Hard X-ray Detector (HXD) instruments, were reduced and analyzed using the standard criteria defined in the ABC Guide\footnote{http://heasarc.gsfc.nasa.gov/docs/suzaku/analysis/abc/}.  These procedures are described below.

\subsubsection{XIS data}
\label{XIS description}
The \textsl{Suzaku} XIS suite consists of four X-ray imaging telescopes each fitted with a CCD chip covering a region 17.$\arcmin$8$\times$17.$\arcmin$8 \citep{2007PASJ...59S..23K, 2007PASJ...59S...1M}.  Three of the chips are front-illuminated -- XIS-0, XIS-2, and XIS-3.  The energy resolution FWHM in the front-illuminated XIS instruments is $\sim$130\,eV at 6.0\,keV.  Sensitive to X-rays ranging from 0.6--10\,keV, the effective area of the front-illuminated XIS instruments is 330\,cm$^2$ at 1.5\,keV.  XIS-2 failed in late 2006 due to a charge leak.  The back-illuminated XIS instrument, XIS-1, is characterized by a greater sensitivity to X-rays in the energy band between 0.2--6\,keV.

4U 1210-64 was observed in full window mode with a data readout of 8\,s.  Data collected using the XIS were reduced and screened using the HEAsoft v.6.13 package and calibration files dated 2013 September 08 (XIS) and 2011 June 30 (XRT) implementing the procedures defined in the \textsl{Suzaku} ABC Guide.  The data were reprocessed with the FTOOL \texttt{aepipeline} using the standard criteria to apply the newest calibration and default screening criteria.  The XIS exposures in the 3$\times$3 and 5$\times$5 event modes were combined using \texttt{XSELECT}.  Circular regions of radius 3.9$\arcmin$ centered on the source and offset from the source were selected to distinguish between photons originating from the source and those originating from the background.  The light curves were binned at 16\,s.

Response matrices were generated using the \texttt{xisrmfgen} and \texttt{xissimarfgen} FTOOL packages.  Pileup was taken into consideration in the region files, where the central pixels were shown to be affected.  As a result, the inner parts of the point spread function (PSF) were removed using two overlapping rectangular shaped regions to reduce pile-up to $\lesssim$1$\%$ using the FTOOLs \texttt{aeattcor2} and \texttt{pileest}.  \texttt{Aeattcor2} creates an improved attitude file, which is then applied to the event file.  The FTOOL \texttt{pileest} was then used on the improved event file, which provides a rough estimate of the degree of pile-up.

Data in the spectral file produced by \texttt{XSELECT} were further processed using the FTOOL \texttt{GRPPHA}.  \texttt{GRPPHA} is designed to define the binning, quality flags and systematic errors of the spectra and used the bad quality flag to further eliminate bad data from the PHA file.  Bins were grouped to ensure a mininum of 20 counts in each in the XIS spectra.  The spectrum was analyzed using \texttt{XSPEC v12.7.1d}.  {\mybf To avoid poorly calibrated Si and Au features, only the energy ranges 0.5--1.7\,keV and 2.1--9.2\,keV were considered} \citep{2011ApJ...728...13N}.  We will discuss the spectral analysis and results in Section 3.4.

\subsubsection{HXD data}

The HXD is a non-imaging X-ray spectrometer consisting of 64 silicon PIN diodes as well as the Gadolinium silicate crystal (GSO) instruments \citep{2007PASJ...59S...1M}.  Our analysis only considers data collected by the PIN diodes because 4U 1210-64 is not bright enough for GSO analysis.  The calibration files used in the analysis of the HXD data were dated 2010 December 6.

The HXD-PIN consists of 16 identical (4$\times$4) detector units surrounded by 20 anti-coincidence counters.  The usable energy range for the HXD PIN diodes is between 15 and 70\,keV \citep{2007PASJ...59S...1M, 2007PASJ...59S..35T}.  The effective area of the HXD instrument is $\sim$160\,cm$^2$ at 20\,keV \citep{2007PASJ...59S...1M, 2007PASJ...59S..35T}.  For this specific observation, 4U 1210-64 is not easily detectable at energies exceeding $\sim$30\,keV.  As a result, the analyzed part of the spectrum is between 15--30\,keV.

The HXD-PIN spectral data were extracted and reduced using the ``cleaned" event files in the \texttt{hxd/event\_cl} directory and the \texttt{hxdpinxbpi} FTOOL package, respectively.  The FTOOL \texttt{hxdpinsbpi} automatically runs the tasks outlined as follows.  Good Time Intervals were calculated using the Non X-ray Background (NXB)\footnote{ftp://legacy.gsfc.nasa.gov/suzaku/data/background/pinnxb\_ver2.0\_tuned/} data overlapping in time with the GTI of the observation.  The source and NXB spectra are extracted using \texttt{hxdpinxbpi}.  The rate in the NXB event file is scaled by a factor of 10 to account for Poisson errors.  As a result, the exposure time of the derived background spectra and light curves must be increased by a factor of 10.  Since the NXB spectrum does not include the Cosmic X-ray Background (CXB), a simulated CXB spectrum was produced using the parameters determined by \citet{1987IAUS..124..611B}.  The total PIN background spectrum is the sum of the NXB and simulated CXB spectra, which were added together using \texttt{addspec}.  The net count rates for the NXB and CXB spectra are 0.2336$\pm$0.0006 and 0.0153$\pm$0.0001\,counts s$^{-1}$, respectively.  The source spectrum is dead time corrected by $\sim$4--5$\%$.

\section{Results}

\subsection{Long Term Temporal Analysis}

\begin{figure}
\centerline{\includegraphics[width=3in]{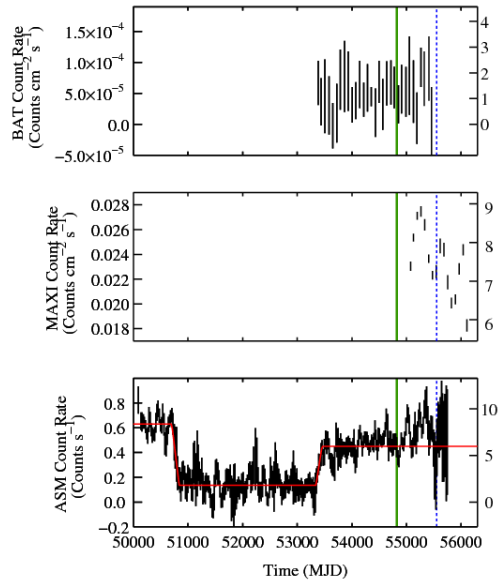}}
\figcaption[2-3-14_rebinned-lightcurves.pdf]{
Long term light curves of 4U 1210-64 produced by BAT in the 15--150\,keV band (top), MAXI in the 2--10\,keV band (middle) and ASM in the 1.5--12\,keV band (bottom) show a two year time overlap between ASM and MAXI and a five\,year time overlap between ASM and BAT.  The ASM, MAXI and BAT light curves use 21\,day, 70\,day and 70\,day time bins, respectively.  {\mybf The times of the PCA} and \textsl{Suzaku} observations are indicated by {\mybf the green shaded region} and blue dashed line, respectively.  The ASM light curve (bottom) is fit with an asymmetric ``step and ramp" function (see solid red line), which models the long-term behavior of the system.
\label{Rebinned Light Curves}
}
\end{figure}

Using data acquired from the ASM, MAXI, and BAT instruments, we produced light curves of 4U 1210-64 to investigate long-term variability of the source (see Figure~\ref{Rebinned Light Curves}).  Data produced by ASM, MAXI and BAT span periods of $\sim$16\,years, $\sim$3.3\,years, and $\sim$5\,years respectively.  

\subsubsection{ASM Temporal Analysis}
\label{ASM Temporal Analysis}

Three distinct states of the system, two active phases and a quiescent phase, are seen in the ASM 
light curve (see Figure~\ref{Rebinned Light Curves}, bottom).  To parameterize the states observed 
in the system, the light curve was fit using an asymmetric ``step-and-ramp" function (see Figure~
\ref{Rebinned Light Curves}, bottom). The parameters in this model are as follows: the times 
corresponding to the start of the transition between the first active state to the quiescent \
state, $T_1$, the start of the transition between the quiescent state and the second active state, 
$T_2$, the transition time between state 1 and state 2, $\Delta T_1$, the transition time between 
state 2 and state 3, $\Delta T_2$, the count rates during both active states, 
$C_{\rm act1}$ and 
$C_{\rm act2}$, and the count rate during the quiescent 
state, $C_{\rm quies}$.  The model parameters 
are reported in Table~\ref{Long Term Behavior}.  The duration of the quiescent phase was found to 
be 2506$^{+36}_{-26}$\,days (6.19$^{+0.10}_{-0.07}$\,yr) using the following equation:

\begin{equation}
\Delta T_{\rm quies}=T_2-(T_1+\Delta T_1)
\end{equation}

\begin{deluxetable}{rrrrr}
\tablecolumns{3}
\tablewidth{0pc}
\tablecaption{Best-fit parameters for the states observed in 4U 1210-64 \label{Long Term Behavior}}
\tablehead{
\colhead{Model Parameter} & \colhead{ASM}}
\startdata
$T_1$ (MJD) & 50703$^{+13}_{-5}$ \\
$\Delta T_1$ (days) & 135$^{+32}_{-22}$ \\
$T_1$+$\Delta T_1$ (MJD) & 50838$^{+35}_{-22}$ \\
$C_{\rm act1}$ (Counts s$^{-1}$) & 0.63$\pm$0.02 \\
$C_{\rm act2}$ (Counts s$^{-1}$) & 0.45$^{+0.02}_{-0.01}$ \\
$T_2$ (MJD) & 53444$^{+11}_{-13}$ \\
$\Delta T_2$ (days) & 109$^{+16}_{-21}$ \\
$T_2$+$\Delta T_2$ (MJD) & 53452$^{+19}_{-25}$ \\
$C_{\rm quies}$ (Counts s$^{-1}$) & 0.136$\pm$0.009 \\
\tableline
$\Delta T_{\rm quies}$ (days) & 2506$^{+36}_{-26}$ \\
\tableline
$\chi^2_\nu$ (dof) & 1.19 (222) \\
\enddata
\label{ASM State Table}
\end{deluxetable}

\begin{figure}
\centerline{\includegraphics[width=3in]{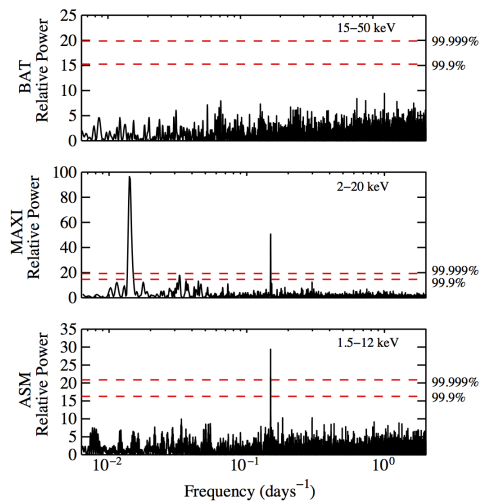}}
\figcaption[6-9-14_power-spectra.pdf]{
Power spectra produced by BAT (top), MAXI (middle) and ASM (bottom) with the 99.9$\%$ and 99.999$\%$ confidence intervals shown.  {\respbf The BAT (top), MAXI (middle) and ASM (bottom) data are in the 15--50\,keV, 2--20\,keV and 1.5-12\,keV bands, respectively.}  The $\sim$70 day precession period of the ISS is also seen in the MAXI power spectrum.
\label{Power Spectrum}
}
\end{figure}

A power spectrum was used to search for periodicities in temporal data \citep{1982ApJ...263..835S, 2008ATel.1861....1C}.  Power spectra, produced using the MAXI and ASM light curves at a time resolution of one dwell (90\,s), show the presence of a 6.7101$\pm$0.0005\,day peak (see Figure~\ref{Power Spectrum}, bottom).  The false-alarm-probability \citep[FAP;][]{1982ApJ...263..835S} in the ASM light curve is $\sim$10$^{-9}$.  We interpret this peak as the orbital period of the system.

\begin{figure}
\centerline{\includegraphics[width=3in]{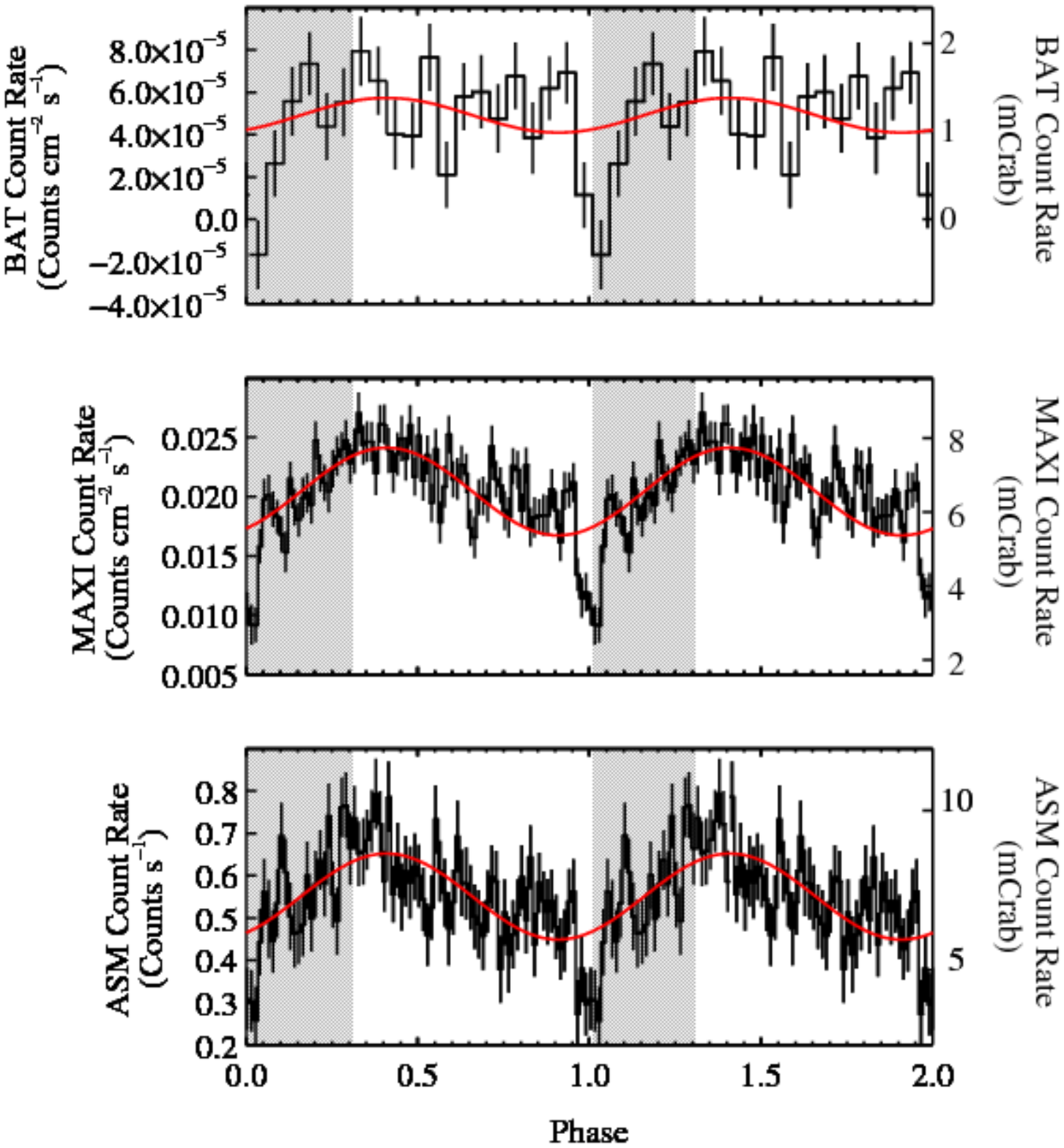}}
\figcaption[8-12-13_sinusoidal-lightcurves.pdf]{
BAT (top), MAXI (middle) and ASM (bottom) lightcurves folded on the orbital period using 20 bins (BAT) and 80 bins (ASM and MAXI).  The shaded region indicates the orbital phases of the system for the duration of the \textsl{Suzaku} observation.  The sharp dip in flux between phases $\phi{\sim}$-0.04 and $\phi{\sim}$0.03 is interpreted as an eclipse.  The solid red line is the sinusoidal fit to the folded light curves using the ephemeris defined in Section 3.1.1.
\label{Folded Light Curves}
}
\end{figure}

{\mybf For this paper, we use an ephemeris based on the orbital period observed in the power spectrum (see Figure~\ref{Folded Light Curves}, bottom) and time of mid-eclipse: $P$=6.7101$\pm$0.0005\,days and $T0$=MJD 54001.8$\pm$0.2.  To parameterize the orbital modulation of the system, the ``dwell-by-dwell" ASM lightcurves were folded using the orbital period.}

\subsubsection{MAXI Temporal Analysis}
\label{MAXI Temporal Analysis}
We confirmed the orbital period using MAXI.  We folded the MAXI light curve over the orbital period using the ephemeris described above (see Figure~\ref{Folded Light Curves}, middle), showing that the binned folded light curve strongly agrees with that produced by ASM.  Compared to the ASM power spectrum, more statistical noise is apparent in the power spectrum produced using MAXI data (see Figure~\ref{Power Spectrum}, middle).  A ``blind search" for the period of the system yields a FAP of $\sim$10$^{-3}$. Considering only the period derived by the ASM data, a single trial search, the FAP is reduced to $\sim$10$^{-6}$.

\subsubsection{BAT Temporal Analysis}

The modulation interpreted as the 6.7 day orbital period was not detected in the power spectrum produced by the BAT instrument (see Figure~\ref{Power Spectrum}, top).  Comparisons between the BAT power spectrum and ASM power spectrum indicate that BAT does not have the sensitivity to detect the 6.7 day modulation at the level seen in ASM and MAXI respectively (see Figure~\ref{Power Spectrum}, top).  We define the modulation depth as ($Count_{\rm max}$$-$$Count_{\rm min}$)$/Mean Count$.

We folded the BAT light curve on the orbital period using the ephemeris defined by the ASM (see Section 3.1.1 and Figure~\ref{Folded Light Curves}, top), showing that the binned folded light curve is consistent with that produced by ASM and MAXI with a possible indication of an eclipse centered at $\phi=$0.

\subsection{Eclipse Profile}
\label{Eclipse Profile}

The folded light curves show the presence of a sharp dip between orbital phases $\phi{\sim}$-0.04 and $\phi{\sim}$0.03, which is suggestive of an eclipse.  {\mybf The source emission does not reach 0\,counts s$^{-1}$.  We interpret this dip as an eclipse since the feature is persistent over many years of data.  The feature is seen at the same orbital phase in States 1 and 3 (see Table~\ref{ASM State Table}), which are separated by $\gtrsim$6\,years.  The rapid ingress and egress requires obscuration by clearly defined boundaries that are suggestive of an object such as the mass donor in the system.}

\begin{figure}
\centerline{\includegraphics[width=3in]{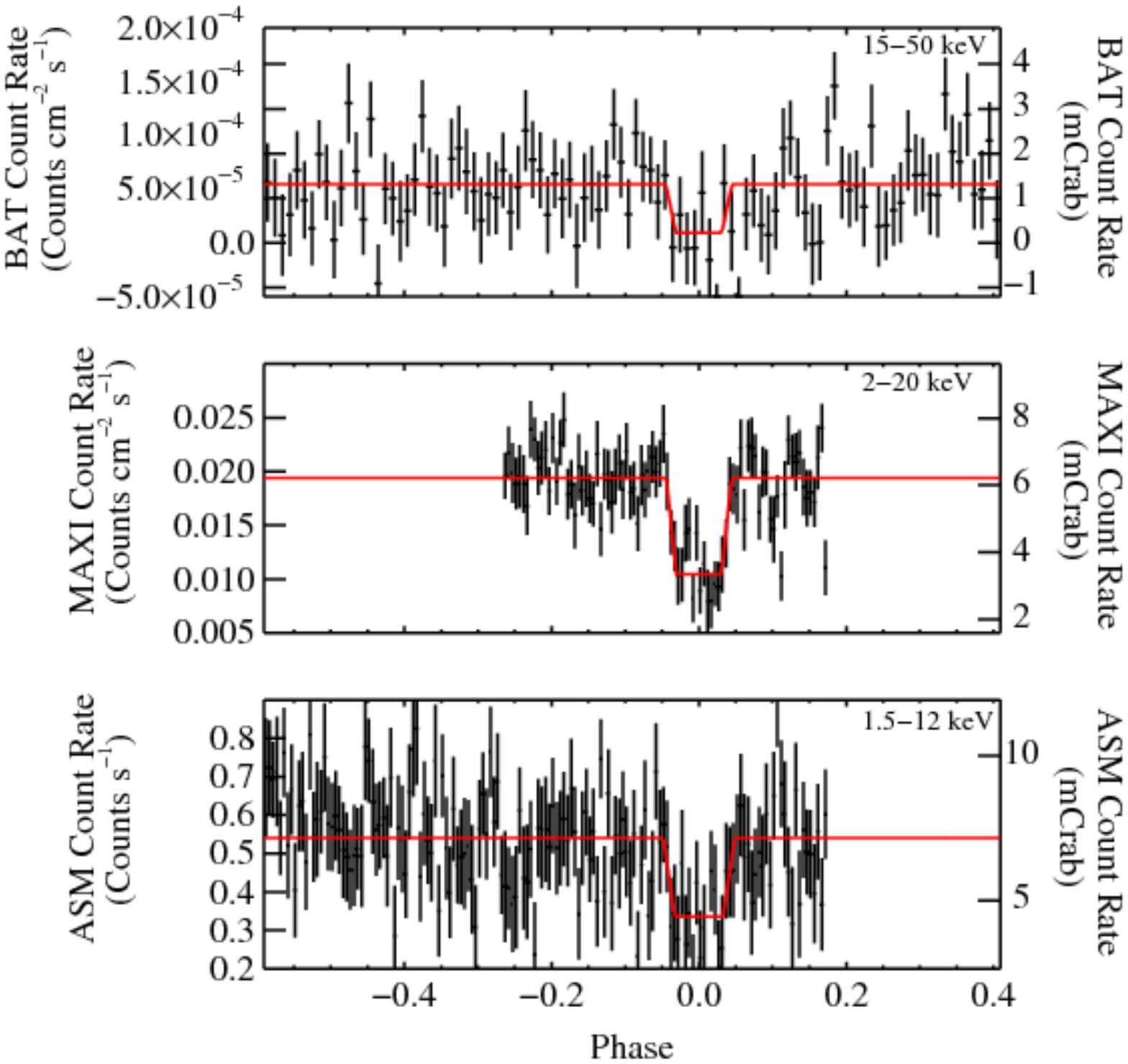}}
\figcaption[6-11-14_eclipse-profile.pdf]{
BAT (top), MAXI (middle) and ASM (bottom) folded light curves are fit with a symmetric ``step and ramp" function, which models the eclipse.  {\respbf The BAT (top), MAXI (middle) and ASM (bottom) data are in the 15--50\,keV, 2--20\,keV and 1.5-12\,keV bands, respectively.}
\label{Eclipse Plot}
}
\end{figure}

The eclipse was modeled using a symmetric ``step and ramp" function (see Figure~\ref{Eclipse Plot}) where the intensities are assumed to remain constant before ingress, during eclipse and after egress and follow a linear trend during the ingress and egress transitions.  The parameters in this model are as follows: the phases corresponding to the start of ingress as well as egress, $\phi_{\rm ing}$ and $\phi_{\rm eg}$, the duration of ingress and egress, $\Delta{\phi}$, the count rates before ingress and after egress, $C$, and the count rate during eclipse, $C_{\rm ecl}$.  The symmetric nature of the model ensures that the duration of ingress and egress, $\Delta{\phi}$, as well as the count rate before ingress and after egress, $C$, are equal.  The eclipse model parameters are reported in Table~\ref{Step and Ramp}.  The eclipse duration and mid-eclipse are calculated using Equations~\ref{Half Angle Equation} and~\ref{Mid Eclipse Equation}.

\begin{equation}
\label{Half Angle Equation}
\Delta{\phi}_{\rm ecl}=\phi_{\rm egr}-(\phi_{\rm ing}+\Delta{\phi})
\end{equation}

\begin{equation}
\label{Mid Eclipse Equation}
\phi_{\rm mid}=\frac{1}{2}(\phi_{\rm egr}+(\phi_{\rm ing}+\Delta{\phi}))
\end{equation}

Since the Suzaku observation begins at MJD 55553.1, we express the time of mid-eclipse, $T_{\rm mid}$, at an epoch closest to the Suzaku observation (see Table~\ref{Step and Ramp}).  The eclipse duration, time of mid-eclipse, and eclipse half-angle ($\Delta{\phi}_{\rm ecl}$$\times$180$\degr$) from fitting the ASM and MAXI folded light curves are reported in Table~\ref{Step and Ramp}.  The fits of both the ASM and MAXI data, which use 150 and 88 bins respectively, indicate that the mid-eclipse times and eclipse durations are in agreement at the 1 $\sigma$ level.  While all model parameters were free for the ASM and MAXI fits, the phases corresponding to the start of ingress as well as egress and the transition duration in the BAT fit, which uses 96 bins, were frozen to the weighted average of the ASM and MAXI values.

\begin{deluxetable}{ccccc}
\tablecolumns{3}
\tablewidth{0pc}
\tablecaption{4U 1210-64 Eclipse Model Parameters}
\tablehead{
\colhead{Model Parameter} & \colhead{ASM} & \colhead{MAXI} & \colhead{Combined} & \colhead{BAT$^a$} \\
\colhead{} & \colhead{(1.5--12\,keV)} & \colhead{(2--20\,keV)} & \colhead{} & \colhead{(15-50\,keV)}}
\startdata
$\phi_{\rm ing}$ & -0.047$^{+0.004}_{-0.003}$ & -0.044$\pm$0.002 & -0.045$\pm$0.002 & \nodata \\
$\Delta{\phi}$ & 0.014$^{+0.001}_{-0.006}$ & 0.013$\pm$0.002 & 0.014$^{+0.001}_{-0.002}$ & \nodata \\
$C$ & 0.54$\pm$0.01$^b$ & 1.97$\pm$0.03$^c$ & \nodata & 5.5$\pm$0.4$^d$ \\
$\phi_{\rm egr}$ & 0.033$\pm$0.005 & 0.031$\pm$0.002 & 0.031$\pm$0.002 & \nodata \\
$C_{\rm ecl}$ & 0.34$^{+0.02}_{-0.03}$$^b$ & 1.11$\pm$0.07$^c$ & \nodata & 1.0$\pm$1.4$^d$ \\
\tableline
$\Delta{\phi}_{\rm ecl}$ & 0.066$^{+0.007}_{-0.008}$ & 0.062$\pm$0.004 & 0.062$\pm$0.003 & \nodata \\
$T_{\rm mid}^e$ & 55553.099$\pm$0.004 & 55553.098$\pm$0.002 & 55553.098$\pm$0.002 & \nodata \\
$\Theta_{\rm e}^f$ & 11.9$^{+1.3}_{-1.5}$ & 11.1$^{+0.6}_{-0.7}$ & 11.2$\pm$0.6 & \nodata \\
\tableline
$\chi^2_\nu$ (dof) & 0.92(146) & 1.21(84) & \nodata & 0.93(96) \\
\enddata
\tablecomments{\\*
$^a$ The $\phi_{{\rm ing}}$, $\Delta{\phi}$, $\phi_{\rm egr}$ parameters in the BAT fit are frozen to the weighted average of the ASM and MAXI values. \\*
$^b$ Units are counts s$^{-1}$. \\*
$^c$ Units are 10$^{-2}$ counts cm$^{-2}$ s$^{-1}$. \\*
$^d$ Units are 10$^{-5}$ counts cm$^{-2}$ s$^{-1}$. \\*
$^e$ Units are MJD. \\*
$^f$ Units are degrees.}
\label{Step and Ramp}
\end{deluxetable}

\subsection{Short Term Temporal Analysis}

\subsubsection{Suzaku}
\label{Suzaku Short Term Temporal Analysis}

\begin{figure}
\centerline{\includegraphics[width=3in]{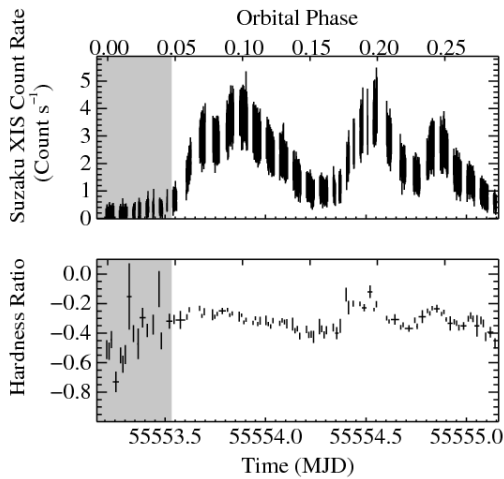}}
\figcaption[12-12-2013_hardness-ratio.pdf]{
The weighted average of the \textsl{Suzaku} XIS0 and XIS3 light curves using bin sizes of 16\,s (top).  The top axis indicates the orbital phase of the system.  The shaded region indicates the egress start and end times calculated by the ``step-and-ramp" function.  The hardness ratio (bottom), which uses bin sizes of 1000\,s, is defined as ($C_{\rm hard}$-$C_{\rm soft}$)/($C_{\rm hard}$+$C_{\rm soft}$), where the soft and hard energy bands are 0.5--4\,keV and 4--10\,keV, respectively.
\label{Suzaku Light Curve}
}
\end{figure}

The two-day \textsl{Suzaku} observation in 2010 Dec. took place during the phase of transition from the minimum to the maximum of the 6.7\,day modulation (see Figure~\ref{Folded Light Curves}). We binned the \textsl{Suzaku} light curves to a resolution of 16\,s to investigate short-term variability in the system, shown in the light curve produced by the sum of the XIS-0 and XIS-3 light curves (see Figure~\ref{Suzaku Light Curve}).

The two-day \textsl{Suzaku} observation revealed large variations in flux indicative of significant variability beyond the orbital modulation.  The modulation depth between the peak in the light curve and the mean count rate is on the order of 140$\%$.

\begin{figure}
\centerline{\includegraphics[width=3in,angle=-90]{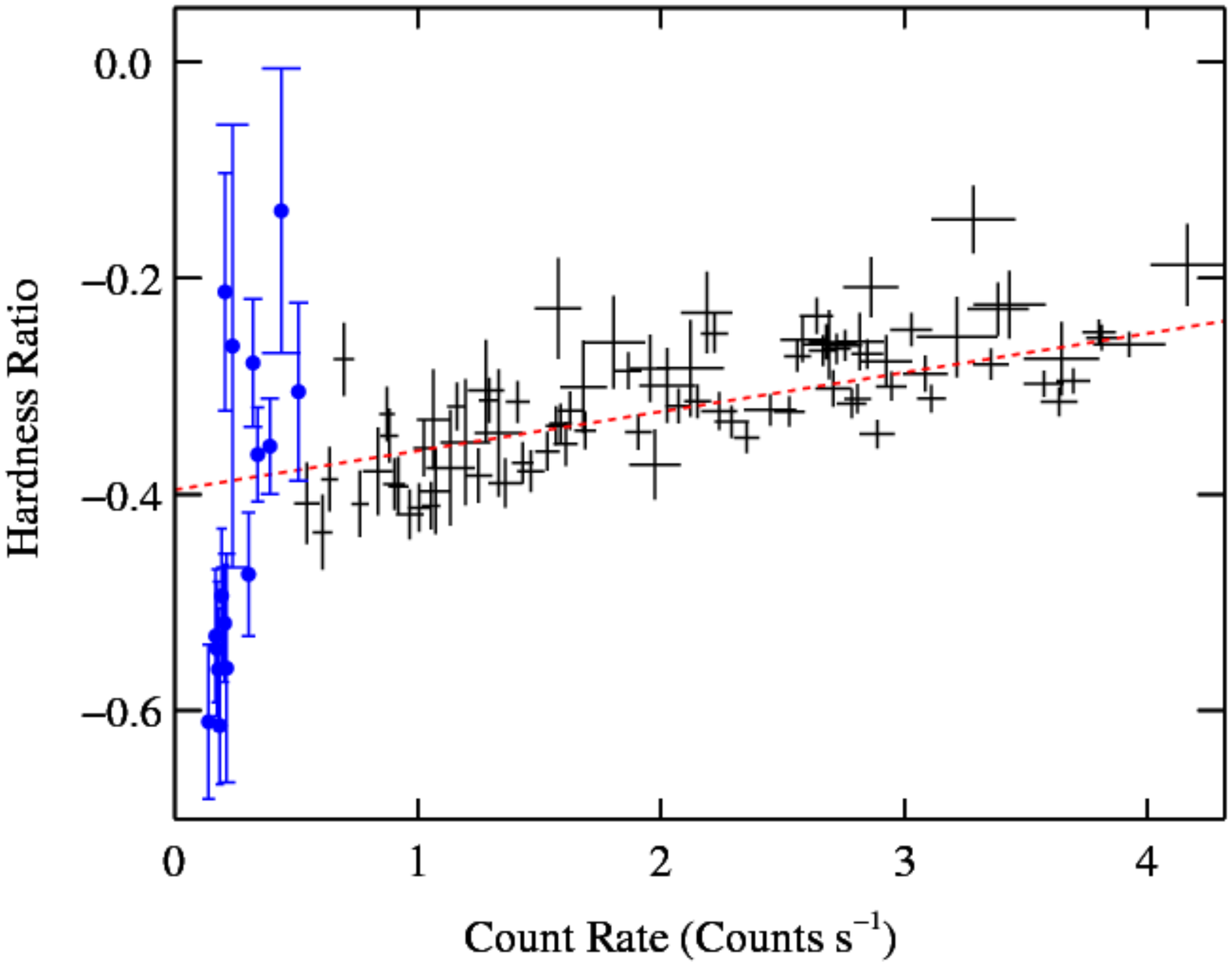}}
\figcaption[february18-2014_hardness_correlation.pdf]{
The hardness ratio as defined in {\mybf Figure 5} vs. the full energy band of 0.5--10\,keV of the \textsl{Suzaku} light curve.  The red dashed line indicates the best linear fit {\mybf to the points outside the eclipse}. {\mybf The blue points indicate the data collected during eclipse.} 
\label{Hardness Correlation}
}
\end{figure}

We divided the light curve into two energy bands where the soft band is defined between energies 0.5--4\,keV, characterized by the count rate $C_{\rm soft}$, and the hard band is between 4--10\,keV, characterized by the count rate $C_{\rm hard}$.  Using the definition in Equation~\ref{Hardness Equation}, we produced a hardness ratio (see Figure~\ref{Suzaku Light Curve}, bottom) binned to a resolution of $\sim$1000\,s, where a soft spectrum is indicated by negative values and a hard spectrum is indicated by positive.  The hardness ratio was also plotted against the count rate in the full energy band of 0.5--10\,keV to search for a correlation between the hardness ratio and source intensity (see Figure~\ref{Hardness Correlation}).  {\respbf Throughout the paper, we use the weighted Pearson correlation coefficient, r \citep[e.g][]{2003drea.book.....B}.}  {\mybf We only take the data observed out-of-eclipse into account since the phenomenology is different from the data observed during eclipse (see Table~\ref{Suzaku Spectral Parameters}).  Out-of-eclipse, we} found a positive correlation between the hardness ratio and source intensity (r=0.69, p$\lesssim$10$^{-6}$), which will be interpreted in Section~\ref{What is the origin of the variability in the high-state?}.

\begin{equation}
\label{Hardness Equation}
HR=(C_{\rm hard}-C_{\rm soft})/(C_{\rm hard}+C_{\rm soft})
\end{equation}

\subsubsection{PCA}
\label{PCA Temporal}

\begin{figure}
\centerline{\includegraphics[width=3in]{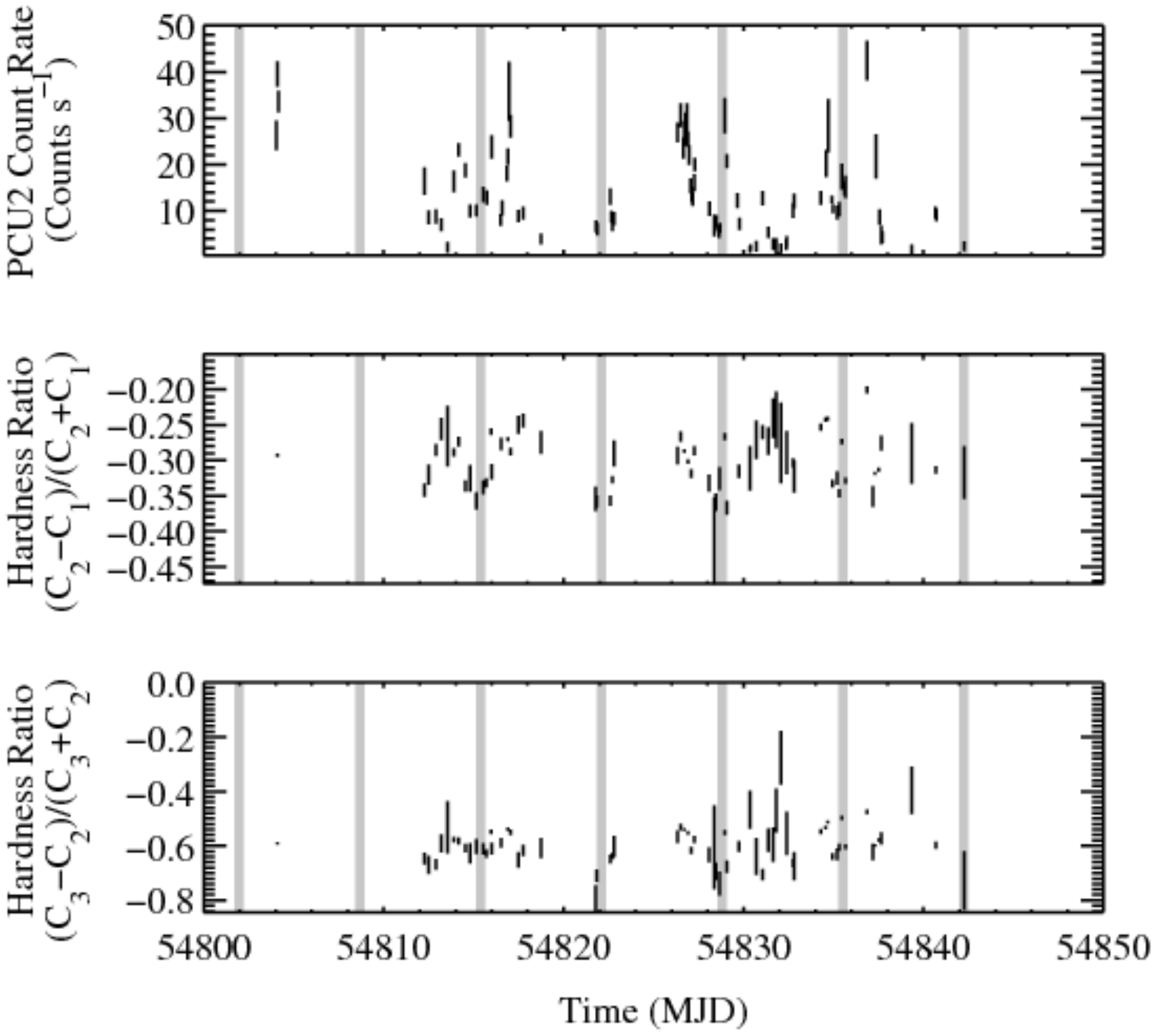}}
\figcaption[may6-2014_PCA_hardness-MJD.pdf]{
Long-term light curve of 4U1210-64 during the RXTE PCA campaign (top).  We calculated the average and standard deviation of the PCU2 count rate for each spacecraft orbit. Also indicated (shaded regions) are the eclipse intervals {\mybf due to the companion.  Hardness ratios of the PCA light curves (middle and bottom), using bin sizes of 0.2\,d., where the energy bands for $C_{\rm 1}$, $C_{\rm 2}$ and $C_{\rm 3}$ are 2.5--6\,keV, 6--10\,keV and 10--20\,keV, respectively.}
\label{PCA Light Curve}
}
\end{figure}

{\mybf To present an overview of the RXTE observations, we calculated the average, background-subtracted count rate of 4U 1210-64 during each spacecraft orbit and plotted this against time in Figure~\ref{PCA Light Curve}.  We divided the PCA light curves into three energy bands defined between between energies 2.5--6\,keV, 6--10\,keV and 10--20\,keV, which are characterized by count rates $C_{\rm 1}$, $C_{\rm 2}$ and $C_{\rm 3}$, respectively.  Using the definition in Equation~\ref{Hardness Equation}, we produced hardness ratios binned to a resolution of $\sim$0.2\,d (see Figure~\ref{PCA Light Curve}, middle and bottom).  To search for correlations between the hardness ratios in each energy band, we produced a color-color diagram (see Figure 8) between the two hardness ratios to examine the correlation between them.  A strong correlation was found in the color-color diagram (r=0.79,p$\lesssim$10$^{-6}$.)}

\begin{figure}
\centerline{\includegraphics[width=3in,angle=-90]{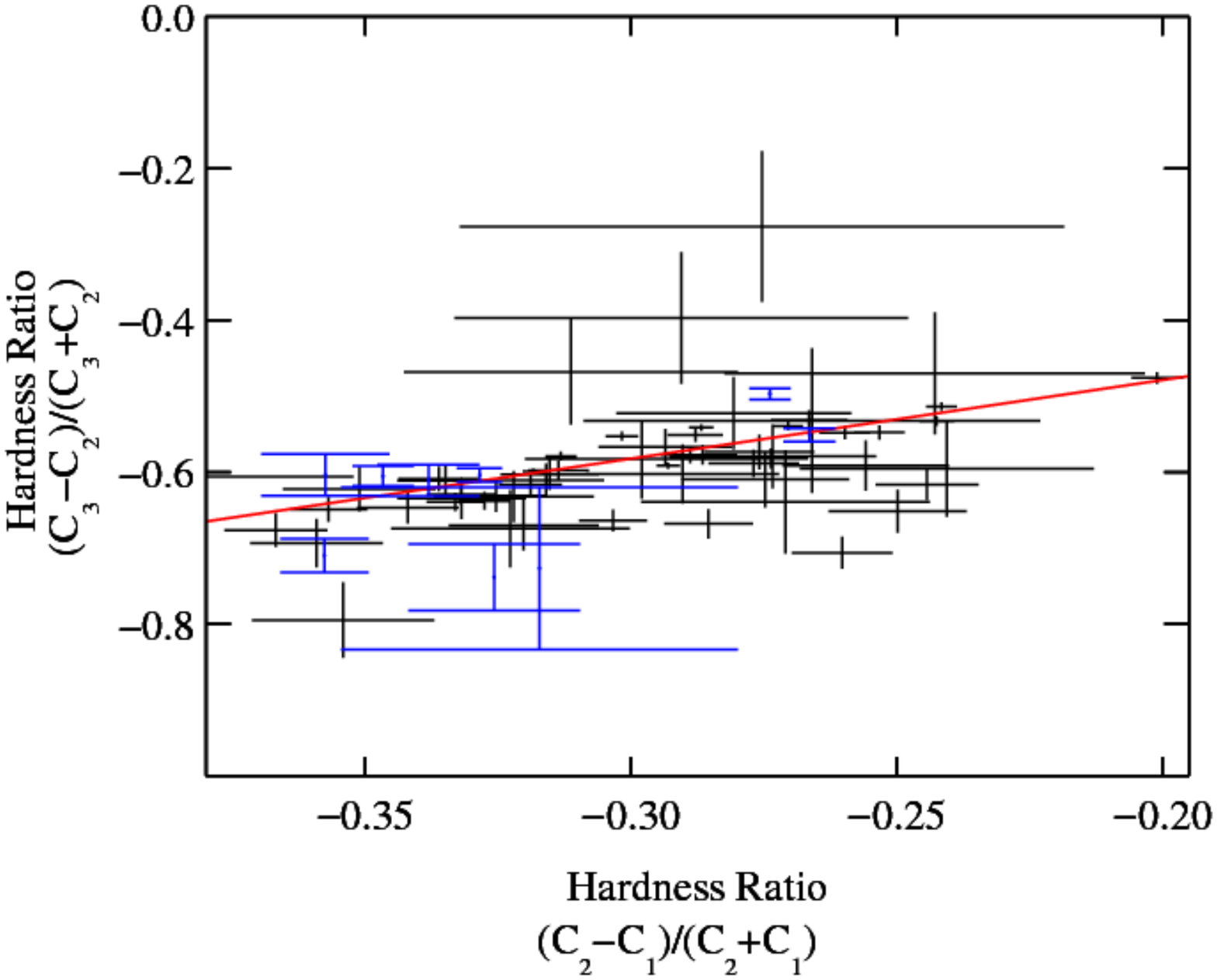}}
\figcaption[may6-2014_PCA-color.pdf]{
PCA color-color diagram comparing soft color vs. hard color as defined in Figure~\ref{PCA Light Curve}.  The red line indicates the best linear fit to the points outside the eclipse. The blue points indicate the data collected during eclipse.
\label{PCA Color Color Diagram}
}
\end{figure}

\begin{figure}
\centerline{\includegraphics[width=3in,angle=-90]{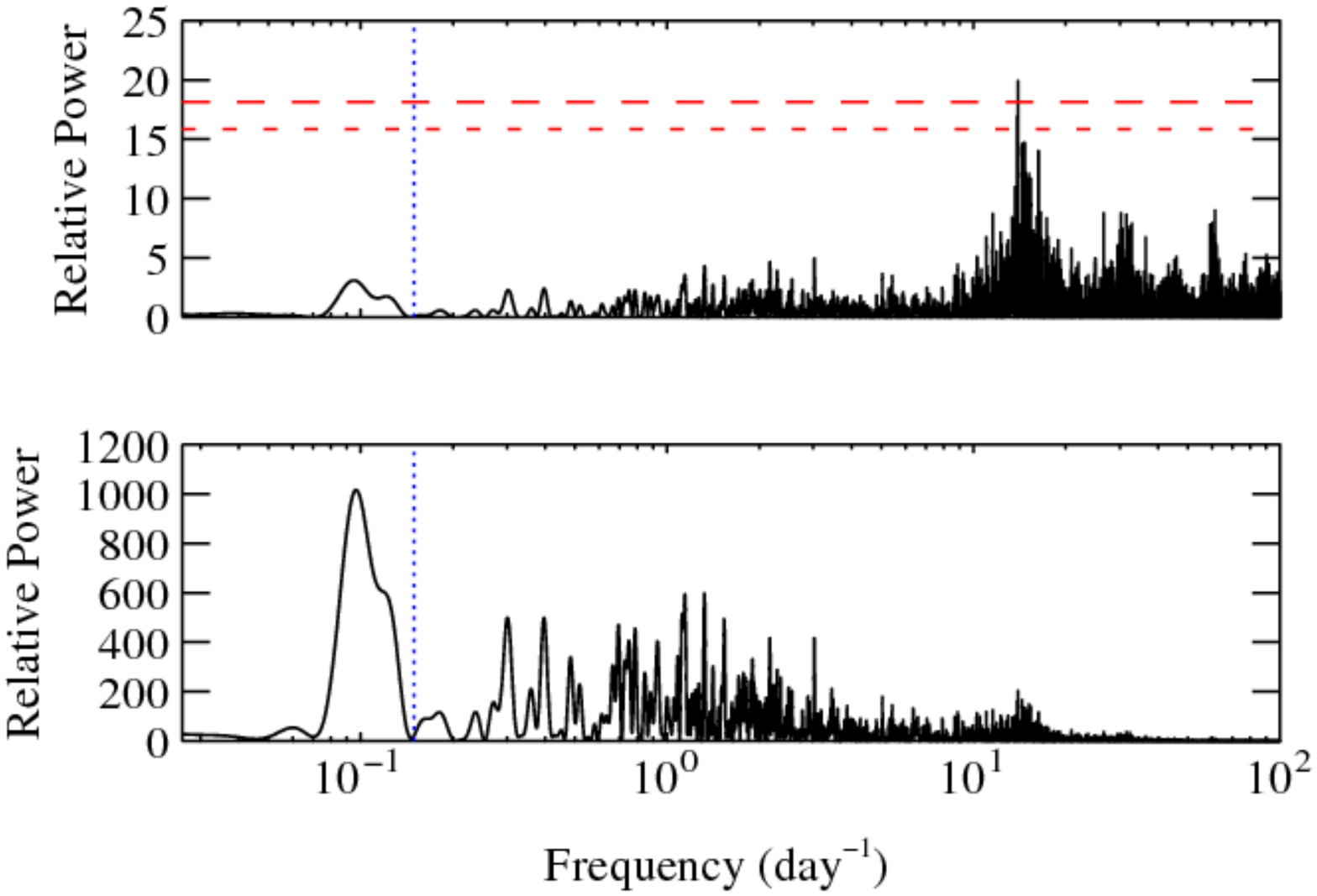}}
\figcaption[4-8-14_PCA_powerspectrum.pdf]{
{\mybf Scargle's power spectrum of the PCA data up to a frequency of 100 cycles per day with the estimated continuum noise component removed (top).  The uncorrected power spectrum (bottom) is dominated by strong low frequency variability.  The orbital period (see Sections~\ref{ASM Temporal Analysis} and~\ref{MAXI Temporal Analysis}) is indicated by the blue dashed line.  The red short dashed line is the power corresponding to 1$\%$ FAP, and the long dashed line 0.1$\%$ FAP.}
\label{Low Frequency}
}
\end{figure}

{\mybf We performed a Fourier transform of all background-subtracted \textsl{RXTE} PCA Standard2 mode data together, up to a frequency of 100\,cycles per day \citep{1982ApJ...263..835S}. The power spectrum is dominated by low frequency (see Figure~\ref{Low Frequency}), ``red" noise, and the orbital period was not detected in this data set.  This is likely because of a patchy orbital phase coverage.}

Note that Scargle's method is strictly correct only in judging the FAP of the highest peak mixed in with otherwise frequency-independent, ``white", noise \citep{1982ApJ...263..835S}.  This is clearly not the case here.  {\mybf To improve our search for a spin period, we analyzed the relationship between power and frequency in log-log space to estimate and remove the amount of low-frequency red noise present in the power spectrum \citep{2005A&A...431..391V}.  A quadratic fit was found to give a reasonable approximation to the continuum noise level.  We subtracted the quadratic fit from the logarithm of the power spectrum along with a constant value of 0.25068 to account for the bias due to the $\chi^2$ distribution of the power spectrum \citep{2005A&A...431..391V}}.  {\respbf The only statistically significant feature in the power spectrum is an artifact caused by a group of peaks around 15 cycles per day near the spacecraft orbital period (see Figure~\ref{Low Frequency}).}

\begin{figure}
\centerline{\includegraphics[width=3in]{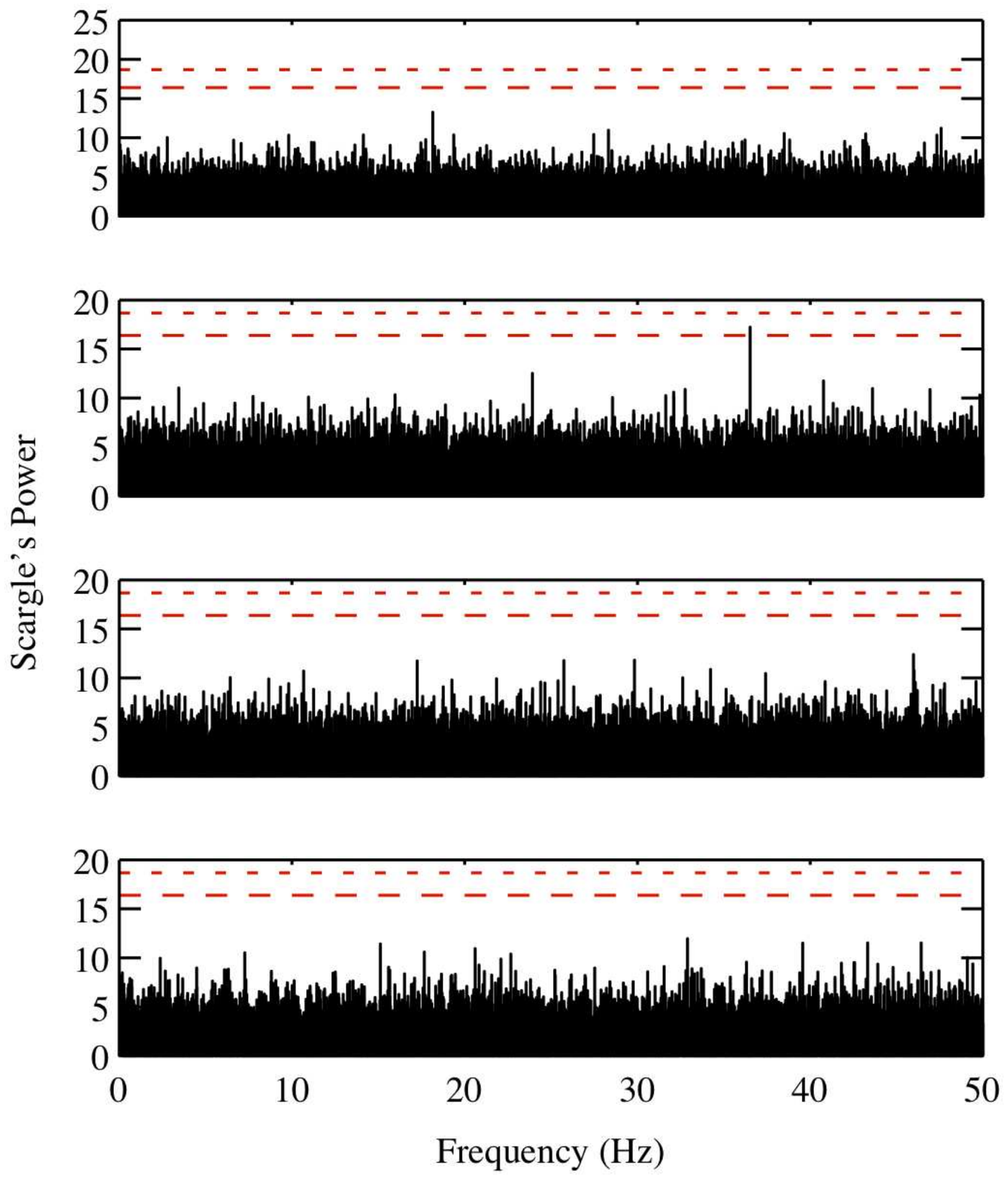}}
\figcaption[april8-2014_Scargle-powerspectra.pdf]{
Typical PCA Power spectra for Obs ID 99009-01-01-17 (top), Obs ID 99009-01-01-22 (second from top), Obs ID 99009-01-23-00 (third panel) and Obs ID 99009-01-03-02 (bottom). {\mybf The frequency range is from 5.73$\times$10$^{-4}$--50\,Hz.}  The long dashed line is the power corresponding to 1$\%$ FAP, and the short dashed line the 0.1$\%$ FAP.  The highest peak in the Obs ID 99009-01-01-22 power spectrum is found to be between the 99$\%$ and 99.9$\%$ confidence intervals (second from top).
\label{Short Power}
}
\end{figure}

We created light curves binned to 10\,ms for each GoodXenon event mode file, covering no more than a single spacecraft orbit.  {\mybf The estimated low-frequency red noise was removed using the procedures in \citet{2005A&A...431..391V}, similar to what we describe above.  A function that is quadratic for frequencies below $\sim$10$^{-2}$\,Hz and constant above this was found to give a reasonable approximation to the continuum noise level found in the log-log plot between power and frequency.}  The highest peak in the resulting power spectra (see Figure~\ref{Short Power}) was often found at the lowest frequency ($<$0.01\,mHz), presumably resulting from both source and background variability on $>$100\,s time scales.  In 5 cases, additional higher frequency ($>$10\,Hz), apparently significant  (FAP less than 0.1$\%$) peaks were also found.  However, these turned out to be related to the low frequency peak: the removal of the low frequency sinusoid also removed these peaks.  We conclude that these were artifacts, created by some (unknown) combination of the sampling pattern, the precise characteristics of the low frequency variability, and possibly also the numerical limitations of the particular implementation of Scargle's algorithm that we used \citep{1982ApJ...263..835S}.  Other than these, the strongest high frequency peaks had FAP between 1$\%$ and 0.1$\%$ (note that the number of trials for each Fourier transform is taken into account, but not the fact that we analyzed 84 independent light curves), and none of these candidate frequencies repeated in multiple observations.  We conclude that we did not detect the spin period of the compact object in 4U 1210-64. Scaling from the amplitudes of the highest peaks, we estimate that a sinusoidal modulation with an amplitude of 8$\%$ of the mean flux would have been detectable at 99.9$\%$ significance.

\subsection{Spectral Analysis}
{\mybf The X-ray spectral data from the \textsl{Suzaku} and PCA observations of 4U 1210-64 were analyzed using the package \texttt{XSPEC v12.8.0}.  We made use of the \texttt{XSPEC} convolution model \texttt{cflux} to calculate the fluxes and associated errors of 4U 1210-64.}

\subsubsection{Suzaku Spectral Analysis}
\label{Suzaku Spectral Analysis}
\begin{figure}
\centerline{\includegraphics[width=2.5in,angle=-90]{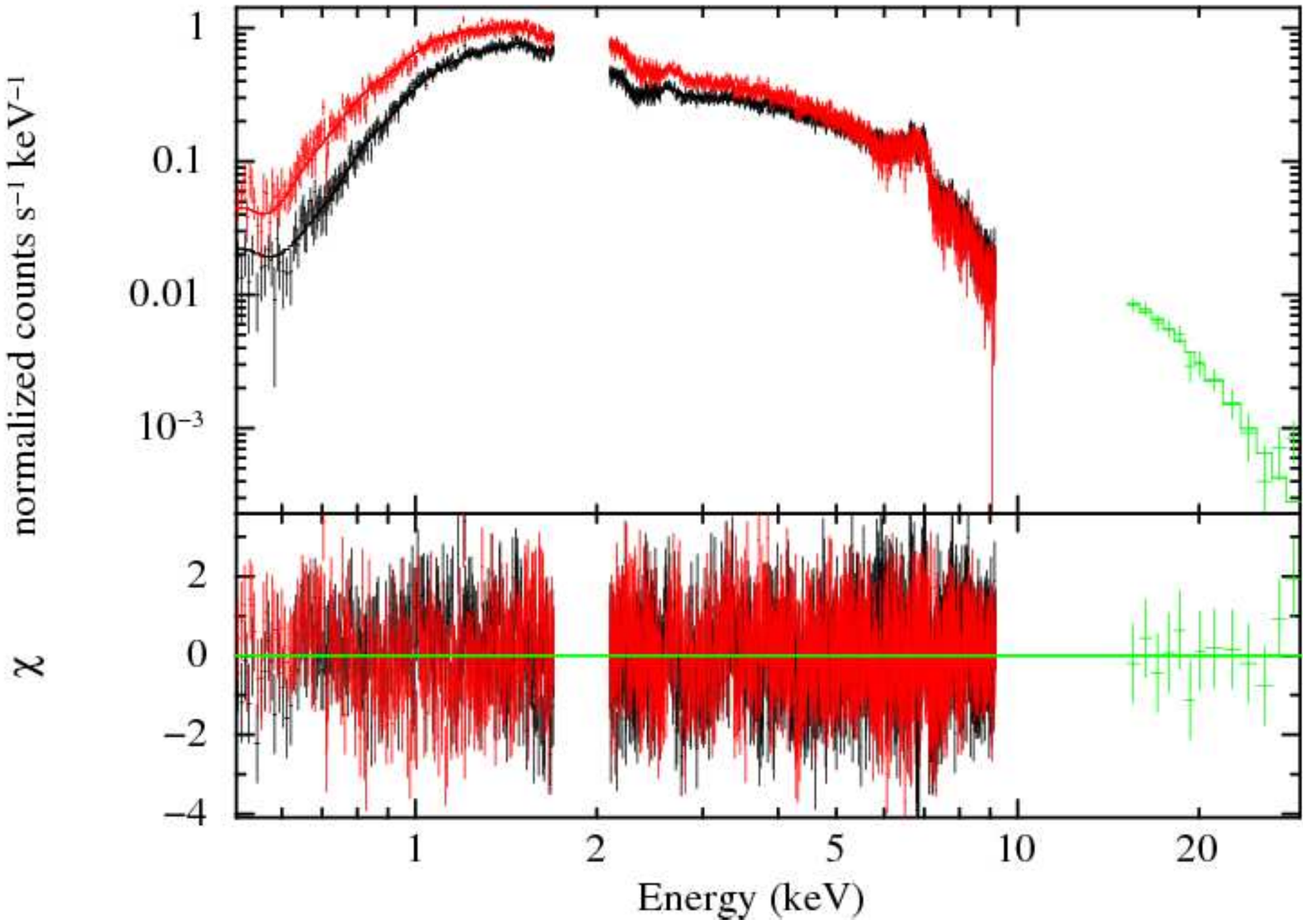}}
\figcaption[11-27-13_cutoffpl-chisq.pdf]{
One of the best fit models for 4U 1210-64 where XIS0$+$3, XIS1 and HXD/PIN are indicated by the black, red and green data/models, respectively.  This consists of a continuum comprised of a power law with a high energy cutoff and four emission lines composed of S XVI K$\alpha$, Fe K$\alpha$ , Fe XXV K$\alpha$ and Fe XXVI K$\alpha$ .  The continuum is absorbed by partial covering absorption and fully covering absorption.
\label{Suzaku Spectra}
}
\end{figure}

To fit the \textsl{Suzaku} spectra, we used several models: a power law, thermal bremsstrahlung, a power law modified with a high energy cutoff (see Figure~\ref{Suzaku Spectra}), and emission due to collisionally-ionized diffuse gas \citep[\texttt{APEC} in \texttt{XSPEC},][]{2012ApJ...756..128F}.  All models were modified by a partially covering absorber in addition to an absorber that fully covers the source using the \citet{1992ApJ...400..699B} cross sections and \citet{2000ApJ...542..914W} abundances.

{\respbf The model that provides a good fit to the data is a power law modified by a high energy cutoff (reported in Table~\ref{Suzaku Spectral Parameters}).  We find that the neutral hydrogen column densities for fully covered and partially covering absorption are $N_{\rm H}$=0.70$\pm$0.01$\times$10$^{22}$ and $N_{\rm H}$=6.7$^{+0.3}_{-0.4}\times$10$^{22}$\,atoms cm$^{-2}$ respectively with a partial covering fraction of 0.36$\pm$0.03.  The measured values of the fully covered absorber are comparable to the Galactic H I values reported by the Leiden/Argentine/Bonn survey \citep{2005A&A...440..775K} and in the review by \citet{1990ARA&A..28..215D}, which are 8.16$\times$10$^{21}$ and 9.37$\times$10$^{21}$\,atoms cm$^{-2}$, respectively.  Therefore, we assume that the fully covered absorber is interstellar in origin unless otherwise noted (see Section~\ref{What is the origin of the variability in the high-state?} for treatment in eclipse).  A good fit does not require an additional blackbody to the power law component, even though such a soft excess was seen in the \textsl{INTEGRAL} data by \citet{2010ApJ...511..A48}.  Furthermore, we do not detect any cyclotron lines in the Suzaku spectra.}

\begin{figure}
\centerline{\includegraphics[width=3.5in]{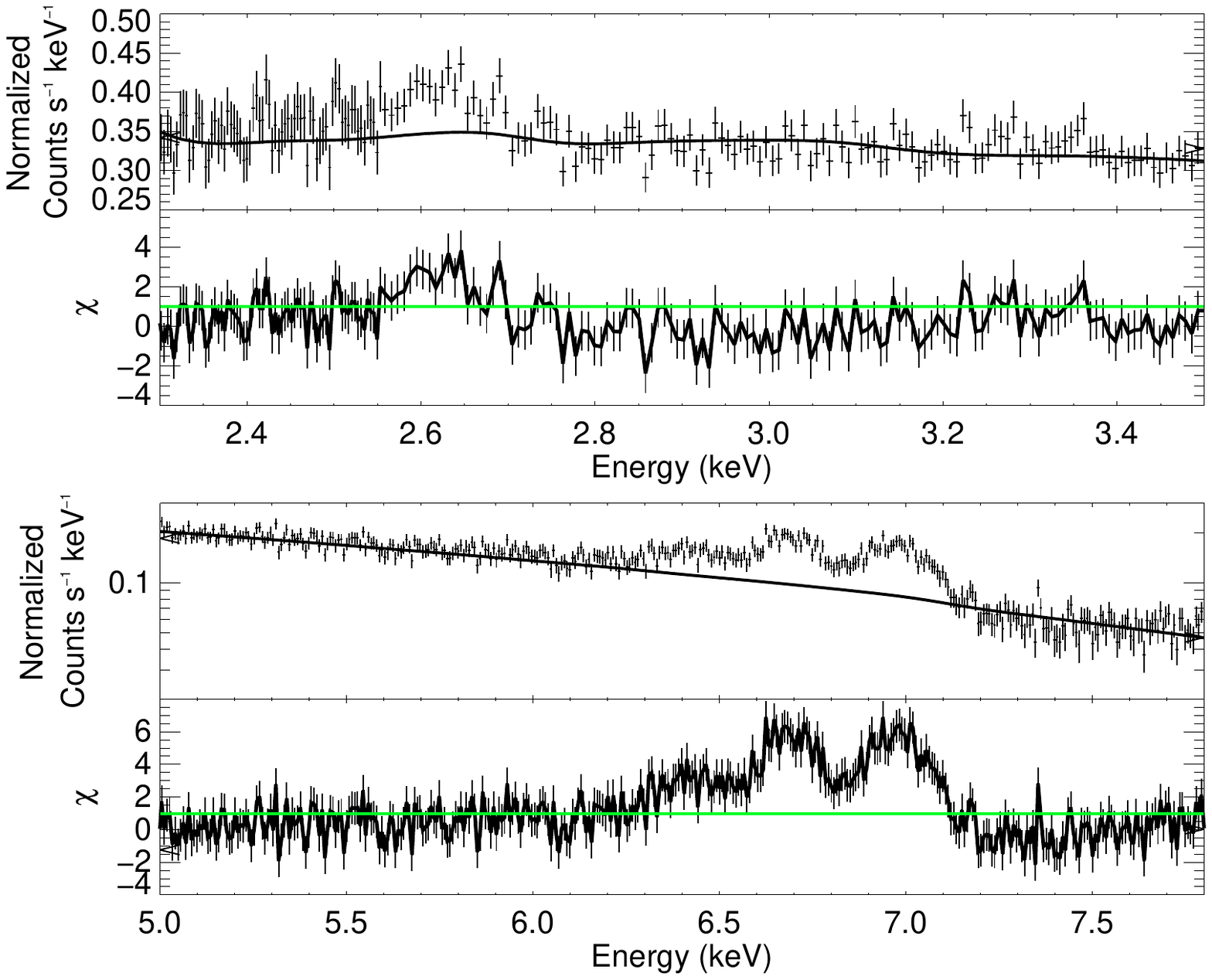}}
\figcaption[august20-2013_Total-Emission-Lines.pdf]{
The XIS-0$+$3 spectrum of 4U 1210-64 in the 2.3--3.5\,keV (top) and 5.0--7.8\,keV bands (third panel) to illustrate the S XVI and Fe K$\alpha$ emission lines along with the best fit model.  The normalization of the lines was set to 0 in the model.  Residuals are plotted in the second from top and bottom panels.
\label{Emission Residuals}
}
\end{figure}

Emission features in the Fe K$\alpha$ region were detected at 6.4\,keV, 6.7\,keV and 6.97\,keV; which were modeled using a Gaussian centered on the peak of the lines (see Figure~\ref{Emission Residuals}).  We interpret these features as Fe K$\alpha$, Fe XXV K$\alpha$ and Fe XXVI K$\alpha$ respectively (see Figure~\ref{Emission Residuals}).  In addition, an emission line was observed at 2.6\,keV, which we interpret as S XVI K$\alpha$ (see Figure~\ref{Emission Residuals}).

\begin{deluxetable}{ccc}
\tablecolumns{3}
\tablewidth{0pc}
\tablecaption{X-ray spectral parameters for 4U 1210-64}
\tablehead{
\colhead{Model Parameter} & \colhead{Cutoff power law} & \colhead{{\respbf Cutoff} power law} \\
\colhead{} & \colhead{{\mybf Out-of-Eclipse}} & \colhead{Eclipse}}
\startdata
$\chi_\nu^2$ (dof) & {\mybf 1.08 (2553)} & 0.81 (382) \\
Cutoff Energy (keV) & {\mybf 5.5$\pm$0.1} & \nodata \\
Folding Energy (keV) & {\mybf 12$\pm$1} & \nodata \\
Phabs $N_{\rm H}$ ($\times$10$^{22}$\,atoms cm$^{-2}$)& {\mybf 0.70$\pm$0.01} & 0.94$\pm$0.08 \\
Pcfabs $N_{\rm H}$ ($\times$10$^{22}$\,atoms cm$^{-2}$)& 6.7$^{+0.3}_{-0.4}$ & 11$^{+2}_{-1}$ \\
Covering Fraction & 0.36$\pm$0.03 & 0.80$^{+0.04}_{-0.05}$ \\
$\Gamma$ & {\mybf 1.80$^{+0.04}_{-0.05}$} & 2.9$\pm$0.2 \\
Normalization ($\times$10$^{-2}$) & {\mybf 2.9$\pm$0.2} & 1.8$^{+0.7}_{-0.5}$ \\
\tableline
S XVI Energy (keV) & {\mybf 2.62$\pm$0.02} & 2.58$\pm$0.05 \\
S XVI Width ($\sigma_{\rm SXVI}$) & {\mybf 0.1$^b$} & 0.1$^b$ \\
Normalization ($\times$10$^{-3}$\,photons cm$^{-2}$ s$^{-1}$) & {\mybf 0.11$\pm$0.02} & 0.11$^{+0.06}_{-0.05}$ \\
S XVI EQW (eV) & {\mybf 20$\pm$3} & 51$^{+26}_{-22}$ \\
S XVI Flux ($\times$10$^{-13}$\,erg cm$^{-2}$ s$^{-1}$) & {\mybf 0.54$\pm$0.09} & 0.4$\pm$0.2 \\
\tableline
Fe I Energy (keV) & {\mybf 6.39$\pm$0.01} & 6.4$^a$ \\
Fe I Width ($\sigma_{\rm Fe I}$) & {\mybf 0.1$^b$} & 0.1$^b$ \\
Normalization ($\times$10$^{-3}$\,photons cm$^{-2}$ s$^{-1}$) & {\mybf 0.093$\pm$0.007} & 0.009$^c$ \\
Fe I EQW (eV) & {\mybf 77$^{+7}_{-5}$} & 122$^c$ \\
Fe I Flux ($\times$10$^{-13}$\,erg cm$^{-2}$ s$^{-1}$) & {\mybf 1.02$\pm$0.08} & 0.7$^c$ \\
\tableline
Fe XXV Energy (keV) & {\mybf 6.684$\pm$0.008} & 6.68$\pm$0.04 \\
Fe XXV Width ($\sigma_{\rm Fe XXV}$) & {\mybf 0.1$^b$} & 0.1$^b$ \\
Normalization ($\times$10$^{-3}$\,photons cm$^{-2}$ s$^{-1}$) & {\mybf 0.185$\pm$0.008} & 0.043$\pm$0.009 \\
Fe XXV EQW (eV) & {\mybf 144$\pm$7} & 392$^{+160}_{-70}$ \\
Fe XXV Flux ($\times$10$^{-13}$\,erg cm$^{-2}$ s$^{-1}$) & {\mybf 1.25$\pm$0.06} & 1.5$\pm$0.3 \\
\tableline
Fe XXVI Energy (keV) & {\mybf 6.970$^{+0.007}_{-0.005}$} & 6.98$\pm$0.05 \\
Fe XXVI Width ($\sigma_{\rm Fe XXVI}$) & {\mybf 0.1$^b$} & 0.1$^b$ \\
Normalization ($\times$10$^{-3}$\,photons cm$^{-2}$ s$^{-1}$) & {\mybf 0.219$^{+0.008}_{-0.009}$} & 0.035$\pm$0.009 \\
Fe XXVI EQW (eV) & {\mybf 199$^{+9}_{-8}$} & 319$^{+160}_{-74}$ \\
Fe XXVI Flux ($\times$10$^{-13}$\,erg cm$^{-2}$ s$^{-1}$) & {\mybf 1.16$\pm$0.05} & 1.5$\pm$0.4 \\
\tableline
Absorbed Flux ($\times$10$^{-10}$\,erg cm$^{-2}$ s$^{-1}$) & {\mybf 1.024$\pm$0.003} & 0.108$\pm$0.004 \\
Unabsorbed Flux ($\times$10$^{-10}$\,erg cm$^{-2}$ s$^{-1}$) & {\mybf 1.73$^{+0.06}_{-0.05}$} & 0.16$\pm$0.01 \\
\enddata
\tablecomments{\\*
$^a$ The energy is frozen because we can only obtain an upper limit. \\*
$^b$ The natural width was frozen to 0.1\,keV, the resolution of the XIS instruments. \\*
$^c$ The upper limit for the parameters (90$\%$ confidence interval) associated with the Fe I line is reported during eclipse.}
\label{Suzaku Spectral Parameters}
\end{deluxetable}

We investigated temporal dependence of the spectral parameters, comparing the parameters during eclipse to those out-of-eclipse. The out-of-eclipse region was subdivided into several 10\,ks intervals to further investigate the temporal variability of the spectral parameters. {\mybf Since 4U 1210-64 is not easily detectable in the HXD-PIN for much of the observation, we only considered the XIS spectra for the analysis of time dependent changes in spectral parameters.  A good fit to the spectra does not require a high-energy cutoff although it is needed for the time-averaged out-of-eclipse spectrum for which we analyzed the well-exposed HXD-PIN spectrum along with the XIS data.  We choose to model the spectra using a power law with a high-energy cutoff frozen at the values of the time-averaged spectra (see Table~\ref{Time Dependent Cutoff Spectrum}).}  The Fe I line was not detected during eclipse (see Table~\ref{Suzaku Spectral Parameters}). We therefore derived an upper limit for the strength of the line.

\begin{deluxetable}{cccccccccccccc}
\rotate
\tablecolumns{14}
\tabletypesize{\small}
\tablewidth{0in}
\tablecaption{Spectral parameters of 4U 1210-64 from the \textsl{Suzaku} observation using the power law continuum with the High Energy Cutoff$^a$}
\tablehead{
\colhead{Time$^b$} & \colhead{Phabs} & \colhead{Pcfabs} & \colhead{Cvr} & \colhead{$\Gamma$} & \colhead{EQW} & \colhead{EQW} & \colhead{EQW} & \colhead{EQW} & \colhead{Flux} & \colhead{$F_{\rm unabs}$} & \colhead{$\chi^2_\nu$} \\
\colhead{(ks)} & \colhead{$N_{\rm H}$} & \colhead{$N_{\rm H}$} & \colhead{} & \colhead{} & \colhead{S XVI} & \colhead{Fe I} & \colhead{Fe XXV} & \colhead{Fe XXVI} & \colhead{Ratio$^c$} & \colhead{(10$^{-10}$ erg cm$^{-2}$ s$^{-1}$)} & \colhead{(d.o.f)} \\
\colhead{} & \colhead{(10$^{22}$ cm$^{-2}$)} & \colhead{(10$^{22}$ cm$^{-2}$)} & \colhead{} & \colhead{} & \colhead{(eV)} & \colhead{(eV)} & \colhead{(eV)} & \colhead{(eV)} & \colhead{} & \colhead{}}
\startdata
0-22.5 & 0.93$\pm$0.08 & 11$^{+2}_{-1}$ & 0.80$^{+0.04}_{-0.05}$ & 2.9$\pm$0.2 & 51$^{+26}_{-22}$ & 70$^d$ & 420$^{+162}_{-76}$ & 364$^{+156}_{-83}$ & 1.0$\pm$0.3 & 0.53$\pm$0.03 & 0.82(382) \\
30-40 & 0.76$\pm$0.06 & 8$\pm$2 & 0.49$^{+0.07}_{-0.08}$ & 1.9$\pm$0.1 & 48$^e$ & 118$^{+50}_{-36}$ & 231$^{+53}_{-40}$ & 147$^{+47}_{-37}$ & 1.3$^{+0.5}_{-0.4}$ & 1.0$\pm$0.1 & 0.92(499) \\
40-50 & 0.71$^{+0.04}_{-0.05}$ & 4$\pm$1 & 0.33$\pm$0.05 & 1.69$^{+0.07}_{-0.06}$ & 16$^e$ & 55$^{+19}_{-20}$ & 100$^{+28}_{-25}$ & 174$^{+39}_{-32}$ & 1.2$\pm$0.3 & 2.1$\pm$0.1 & 0.96(1146) \\
50-60 & 0.69$\pm$0.03 & 5.7$\pm$0.9 & 0.38$\pm$0.04 & 1.76$\pm$0.05 & 24$^{+7}_{-8}$ & 67$^{+14}_{-13}$ & 100$^{+18}_{-15}$ & 171$^{+21}_{-19}$ & 1.0$\pm$0.2 & 2.5$\pm$0.1 & 0.97(2030) \\
60-70 & 0.75$\pm$0.02 & 6.3$\pm$0.7 & 0.43$\pm$0.04 & 1.94$\pm$0.05 & 18$\pm$8 & 65$^{+16}_{-13}$ & 121$^{+19}_{-16}$ & 193$^{+23}_{-19}$ & 1.0$\pm$0.2 & 3.0$^{+0.2}_{-0.1}$ & 0.92(2094) \\
70-80 & 0.76$\pm$0.03 & 6.6$\pm$0.9 & 0.41$\pm$0.04 & 1.97$\pm$0.06 & 18$\pm$9 & 50$^{+17}_{-15}$ & 205$^{+26}_{-21}$ & 206$^{+27}_{-25}$ & 1.2$\pm$0.2 & 2.1$\pm$0.1 & 0.97(1572) \\
80-90 & 0.68$^{+0.04}_{-0.05}$ & 5$\pm$2 & 0.32$\pm$0.07 & 1.84$\pm$0.09 & 23$^{+12}_{-13}$ & 59$^{+26}_{-22}$ & 158$^{+33}_{-26}$ & 263$^{+47}_{-39}$ & 1.2$\pm$0.3 & 1.3$\pm$0.1 & 0.96(1103) \\
90-100 & 0.72$^f$ & 6$\pm$1 & 0.35$^{+0.08}_{-0.09}$ & 1.95$^{+0.08}_{-0.09}$ & 42$^{+16}_{-15}$ & 129$^{+31}_{-29}$ & 203$^{+37}_{-31}$ & 264$^{+46}_{-41}$ & 1.0$^{+0.3}_{-0.2}$ & 0.82$^{+0.07}_{-0.06}$ & 0.94(809) \\
100-110 & 0.67$\pm$0.05 & 12$\pm$3 & 0.3$\pm$0.1 & 1.6$\pm$0.1 & 26$^{+17}_{-18}$ & 108$^{+33}_{-27}$ & 207$^{+46}_{-31}$ & 174$^{+45}_{-31}$ & 1.1$^{+0.4}_{-0.3}$ & 1.2$\pm$0.1 & 0.91(631) \\
110-120 & 0.70$\pm$0.04 & 9$\pm$1 & 0.42$^{+0.06}_{-0.07}$ & 1.69$\pm$0.08 & 15$^{+12}_{-13}$ & 88$^{+23}_{-20}$ & 110$^{+25}_{-20}$ & 168$^{+32}_{-27}$ & 1.2$\pm$0.3 & 2.8$\pm$0.2 & 0.96(1022) \\
120-130 & 0.66$^{+0.04}_{-0.05}$ & 5$\pm$2 & 0.27$\pm$0.07 & 1.73$\pm$0.08 & 20$^{+11}_{-12}$ & 31$^{+21}_{-20}$ & 155$^{+31}_{-25}$ & 215$^{+38}_{-30}$ & 1.0$\pm$0.3 & 2.1$^{+0.2}_{-0.1}$ & 0.96(1021) \\
130-140 & 0.70$\pm$0.03 & 9$\pm$1 & 0.37$\pm$0.07 & 1.86$^{+0.09}_{-0.08}$ & 38$^{+13}_{-12}$ & 26$^{+21}_{-20}$ & 247$^{+37}_{-28}$ & 206$^{+34}_{-25}$ & 1.0$^{+0.3}_{-0.2}$ & 1.4$\pm$0.1 & 1.07(1036) \\
140-150 & 0.66$\pm$0.03 & 7$\pm$1 & 0.30$\pm$0.06 & 1.66$\pm$0.06 & 17$^e$ & 101$^{+18}_{-17}$ & 126$^{+21}_{-16}$ & 190$^{+24}_{-23}$ & 1.1$\pm$0.2 & 1.9$\pm$0.1 & 0.95(1616) \\
150-160 & 0.72$\pm$0.03 & 7$\pm$1 & 0.40$^{+0.06}_{-0.07}$ & 1.89$\pm$0.08 & 16$\pm$12 & 78$^{+23}_{-20}$ & 198$^{+34}_{-25}$ & 231$^{+33}_{-28}$ & 1.1$^{+0.3}_{-0.2}$ & 1.4$\pm$0.1 & 0.98(1108) \\
160-170 & 0.63$^f$ & 16$^{+13}_{-5}$ & 0.23$^{+0.08}_{-0.09}$ & 1.64$^{+0.08}_{-0.07}$ & 61$\pm$16 & 161$^{+34}_{-30}$ & 164$^{+32}_{-27}$ & 242$^{+44}_{-36}$ & 1.1$\pm$0.3 & 0.70$^{+0.05}_{-0.04}$ & 0.99(786) \\
\tableline
\enddata
\tablecomments{The best-fit parameters for the eclipse and out-of-eclipse, which is subdivided into 10 ks time intervals. \\*
$^a$ The cutoff and folding energies of the high energy cutoff are frozen to the best fit values of the out-of-eclipse spectrum: 5.5\,keV and 12.0\,keV, respectively. \\*
$^b$ Time is relative to the start of the \textsl{Suzaku} observation MJD 55553.16. \\*
{\respbf $^c$ The ratio of the strengths of the Fe XXV line with respect to the Fe XXVI line.} \\*
$^d$ The upper limit for the EQW of the Fe I line at the 90$\%$ confidence interval is reported during eclipse. \\*
$^e$ The upper limit for the EQW of the S XVI line at the 90$\%$ confidence interval is reported at the time intervals 30--40\,ks, 40--50\,ks, and 140--150\,ks. \\*
$^f$ The fully covered absorber is frozen to the value that was extracted using the power law model.}
\label{Time Dependent Cutoff Spectrum}
\end{deluxetable}

\begin{figure}
\centerline{\includegraphics[width=3.5in]{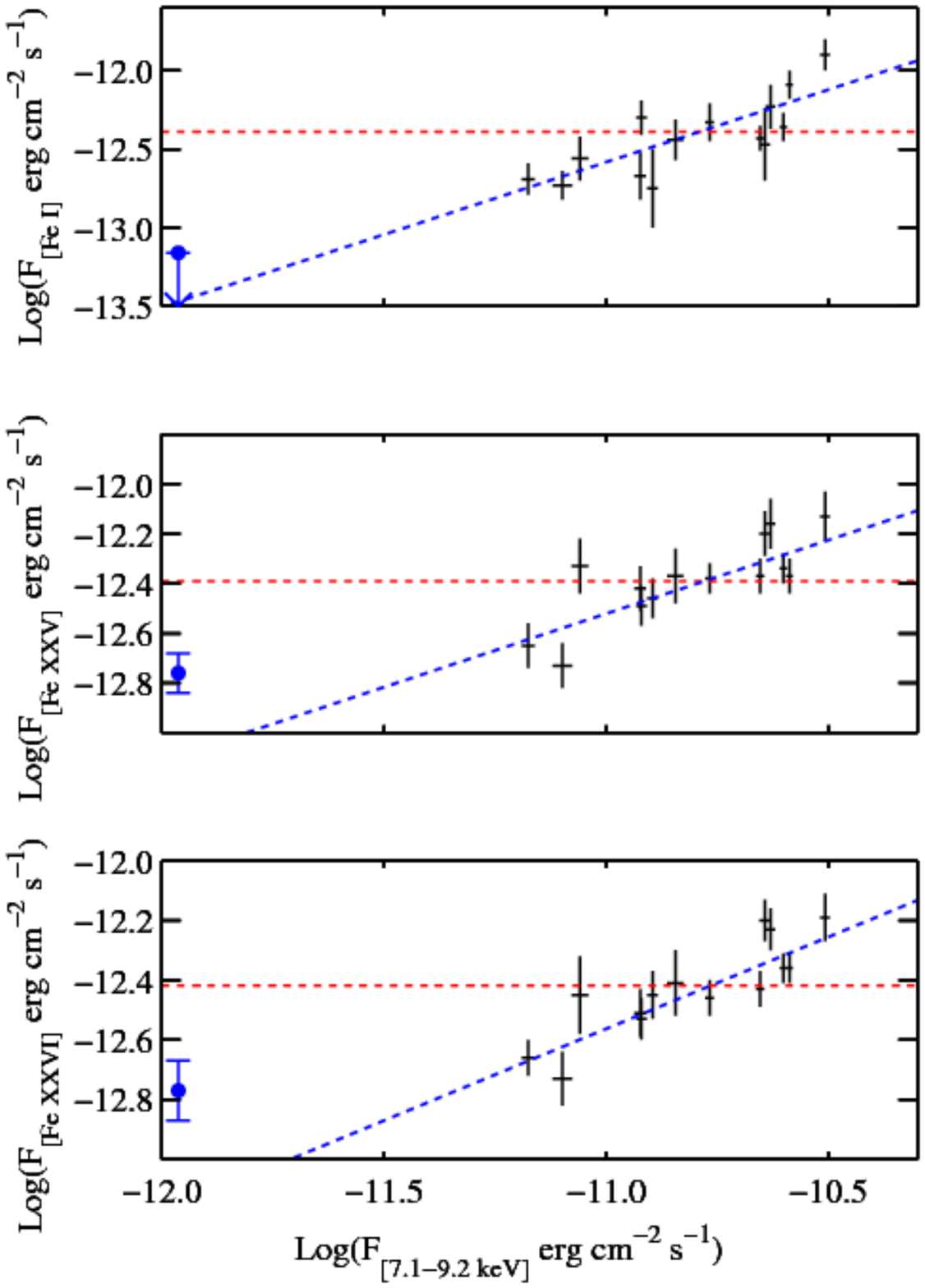}}
\figcaption[3-13-14_highecut_FeKalpha-flux.pdf]{
Flux of the Fe I (top), Fe XXV (middle) and Fe XXVI (bottom) lines vs. the unabsorbed continuum flux in the 7.1--9.0\,keV band in logarithmic units. An upper limit at the 90$\%$ confidence interval is shown for the Fe I line during eclipse.  The red and blue lines indicate the best fit for constant and power law models respectively.  The correlation coefficients, r, are as follows: r=0.82, r=0.81 and r=0.87 for the Fe I, Fe XXV and Fe XXVI lines, respectively.  Using a logarithmic parameter space, we measured the slope of the Fe K$\alpha$ line flux versus continuum flux--m=1.0$\pm$0.1, m=0.6$\pm$0.1 and m=0.7$\pm$0.1 for the Fe I, Fe XXV and Fe XXVI lines, respectively. {\mybf The blue points indicate the data collected during eclipse.}
\label{Flux Flux}
}
\end{figure}

In order to place constraints on the origin of the Fe K$\alpha$ lines, we compared the flux of the 
Fe K$\alpha$ lines with the unabsorbed continuum flux in the 7.1--9.0\,keV band (see Figure~\ref
{Flux Flux}).  The flux of the Fe I line is found to decrease by a factor of $\sim$15 during 
eclipse (see Table~\ref{Time Dependent Cutoff Spectrum}).  Additionally, we found that the flux of 
Fe XXV and Fe XXVI decreases by a factor of approximately 3 during the eclipse phase (see Table~
\ref{Time Dependent Cutoff Spectrum}).  Within measurement errors, the flux of the Fe K$\alpha$ 
lines {\respbf observed out-of-eclipse} was found to follow the flux of the continuum.  The 
correlation coefficients, r, are as follows: r=0.82, r=0.81 and r=0.87 for the Fe I, Fe XXV and Fe 
XXVI lines, respectively.  Using a logarithmic parameter space, we measured the slope of the Fe K$
\alpha$ line flux {\respbf observed out-of-eclipse} versus continuum flux--m=1.0$\pm$0.1, m=0.6$\pm
$0.1 and m=0.7$\pm$0.1 for the Fe I, Fe XXV and Fe XXVI lines, respectively.

{\respbf To constrain the state of the plasma, we calculated the flux ratio between the Fe XXV and Fe XXVI emission features.  We defined the flux ratio as the flux of the Fe XXV line divided by the flux of the Fe XXVI line.  No change was found in the flux ratio between the Fe XXV and Fe XXVI emission features and the continuum flux in the 7.1--9.0\,keV band (see Table~\ref{Time Dependent Cutoff Spectrum}).}

In contrast to the Fe K$\alpha$ lines, the S XVI K$\alpha$ line was not consistently detected throughout the duration of the \textsl{Suzaku} observation (see Table~\ref{Time Dependent Cutoff Spectrum}).  During the time intervals specified in Table~\ref{Time Dependent Cutoff Spectrum}, we calculated an upper limit of the EQW and flux of the S XVI line at the 90$\%$ confidence interval.  We also searched for correlations between the continuum spectral parameters (the power law {\respbf with the high energy cutoff} modified with fully covered and partial covering absorption) with respect to the 7.1--9.0\,keV continuum flux (see Table~\ref{Time Dependent Cutoff Spectrum}).  We found no clear correlation between the parameter values and flux.

{\respbf In addition, we found that a bremsstrahlung model {\newbf with a temperature of 7.2$\pm$0.2\,keV} also fits the \textsl{Suzaku} data reasonably well. However, this model is found to provide unsatisfactory fits to the PCA spectra where $\chi^2_\nu$ was found to exceed 2 (see Section~\ref{PCA Spectral Description}).  Therefore, the power law with the high energy cutoff model is the preferred description of 4U 1210-64.}

\subsubsection{PCA Spectral Analysis}
\label{PCA Spectral Description}

\begin{figure}
\centerline{\includegraphics[width=3.5in]{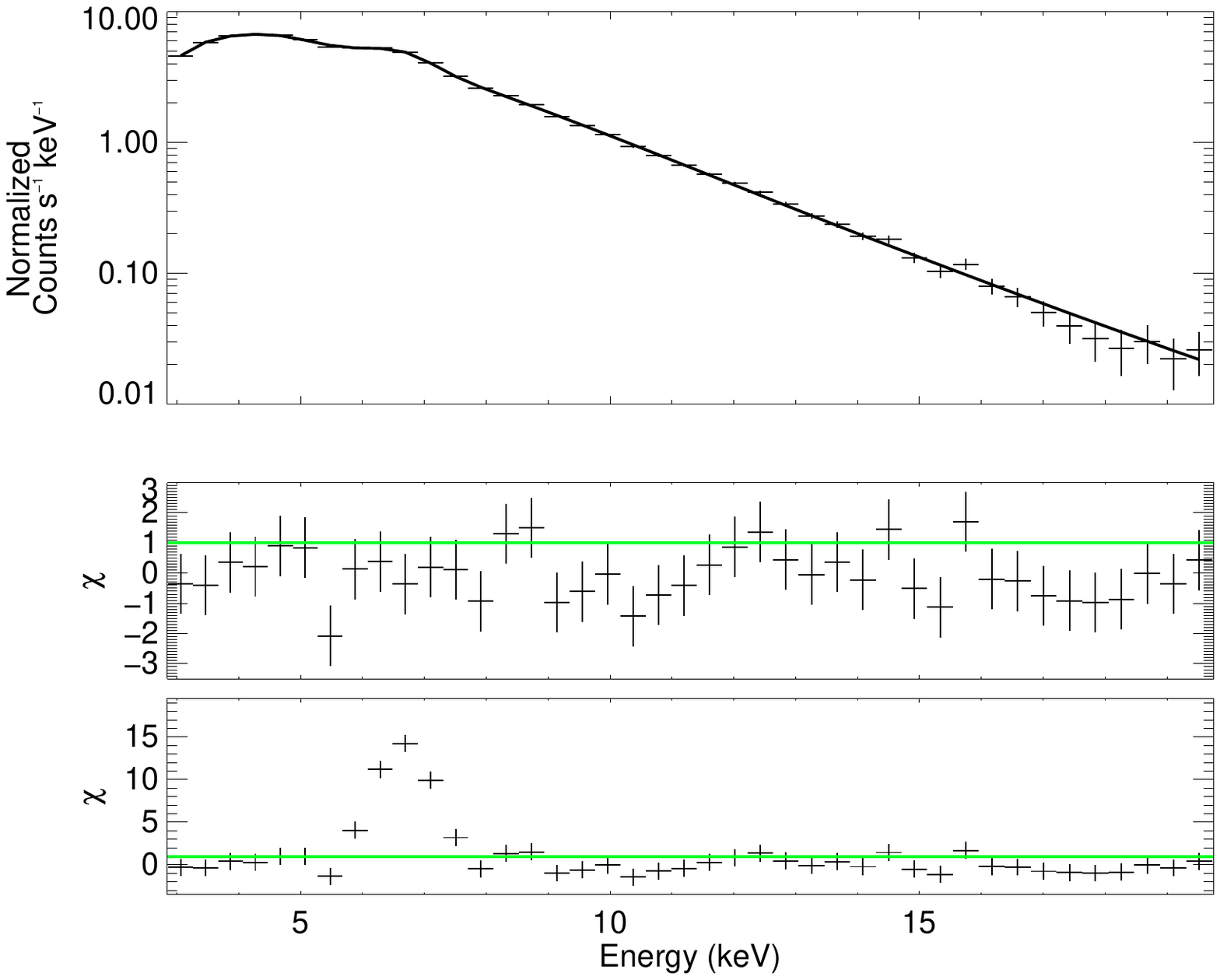}}
\figcaption[6-13-14_Total-PCAspectrum.pdf]{
{\respbf A typical PCA spectrum (MJD 54804) and best fit model (top panel).  This consists of a continuum comprised of a power law with a high energy cutoff and a broad Fe K$\alpha$ emission line.  Residuals of the best fit model are plotted in the middle panel.  To illustrate the Fe K$\alpha$ emission lines along with the best fit model,  the normalization of the line was set to 0.  Residuals are plotted in the bottom panel.}
\label{PCA Spectra}
}
\end{figure}

We analyzed the PCA spectral data of 4U 1210-64 from MJD 54804--54842, using the models described in Section~\ref{Suzaku Spectral Analysis}. The best-fit model is a power law with a high energy cutoff (see Figure~\ref{PCA Spectra}, top and middle).  Due to the lack of low energy response (see Section~\ref{PCA Description}), {\respbf the absorption that is observed in the \textsl{Suzaku} spectra cannot be accurately measured using the PCA (see Section~\ref{Suzaku Spectral Analysis}).  Therefore, we froze $N_{\rm H}$ to the out-of-eclipse value determined with \textsl{Suzaku}.}

{\respbf The power law index, $\Gamma$, and cutoff energies found with the PCA are consistent with those determined with the \textsl{Suzaku} analysis.  However, we note that the folding energy obtained with the PCA differs somewhat from that obtained with \textsl{Suzaku}.  We also note that the folding energy was found to be variable among different time segments of PCA data.}

{\mybf Due to the $\sim$10 times lower spectral resolution of the PCA compared with \textsl{Suzaku} (see Sections~\ref{PCA Description} and~\ref{XIS description}, respectively), we detected only a broad emission feature between energies 6.4--6.7\,keV in the Fe K$\alpha$ region, which was modeled using a single Gaussian (see Figure~\ref{PCA Spectra}, bottom).  Furthermore, the S XVI K$\alpha$ feature was not detected in the PCA spectra, which is expected due to the low sensitivity of the PCA at 2.62\,keV.}

\begin{deluxetable}{cccccccccccc}
\rotate
\tablecolumns{12}
\tabletypesize{\scriptsize}
\tablewidth{10.7in}
\tablecaption{Spectral parameters of 4U 1210-64 from the PCA observations}
\tablehead{
\colhead{Time} & \colhead{Phabs} & \colhead{Pcfabs} & \colhead{Cvr} & \colhead{High Energy} & \colhead{Folding} & \colhead{Power Law} & \colhead{Power Law} & \colhead{Fe K$\alpha$} & \colhead{EQW} & \colhead{$F_{\rm unabs}$} & \colhead{$\chi^2_\nu$} \\
\colhead{(MJD)} & \colhead{$N_{\rm H}$} & \colhead{$N_{\rm H}$} & \colhead{} & \colhead{Cutoff} & \colhead{Energy} &\colhead{$\Gamma$} & \colhead{Normalization} & \colhead{Energy} & \colhead{Fe K$\alpha$} & \colhead{(10$^{-10}$ erg cm$^{-2}$ s$^{-1}$)} & \colhead{(d.o.f)} \\
\colhead{} & \colhead{(10$^{22}$ cm$^{-2}$)} & \colhead{(10$^{22}$ cm$^{-2}$)} & \colhead{} & \colhead{(keV)} & \colhead{(keV)} & \colhead{} & \colhead{(10$^{-2}$)} & \colhead{(keV)} & \colhead{(eV)} & \colhead{} & \colhead{}}
\startdata
54804 & 0.70$^a$ & 6.74$^a$ & 0.36$^a$ & 6.0$\pm$0.1 & 5.6$\pm$0.1 & 1.50$\pm$0.03 & 7.8$\pm$0.3 & 6.66$^{+0.01}_{-0.02}$ & 228$^{+26}_{-18}$ & 7.4$\pm$0.2 & 0.82(35) \\
54812 & 0.70$^a$ & 6.74$^a$ & 0.36$^a$ & 7.4$\pm$0.2 & 5.5$\pm$0.4 & 1.96$\pm$0.04 & 4.6$\pm$0.3 & 6.65$^{+0.01}_{-0.07}$ & 414$^{+52}_{-39}$ & 1.71$\pm$0.04 & 0.49(35) \\
54813 & 0.70$^a$ & 6.74$^a$ & 0.36$^a$ & 6.6$\pm$0.3 & 7.0$^{+0.5}_{-0.4}$ & 1.70$\pm$0.05 & 3.2$\pm$0.2 & 6.59$^{+0.08}_{-0.02}$ & 306$^{+30}_{-36}$ & 2.01$\pm$0.06 & 0.71(35) \\
54814 & 0.70$^a$ & 6.74$^a$ & 0.36$^a$ & 6.5$\pm$0.2 & 5.9$\pm$0.3 & 1.70$\pm$0.04 & 4.6$^{+0.3}_{-0.2}$ & 6.66$\pm$0.09 & 159$^{+27}_{-32}$ & 2.86$\pm$0.08 & 0.35(35) \\
54815 & 0.70$^a$ & 6.74$^a$ & 0.36$^a$ & 6.3$\pm$0.2 & 6.8$^{+0.2}_{-0.3}$ & 1.70$\pm$0.03 & 5.0$\pm$0.2 & 6.662$^{+0.004}_{-0.085}$ & 250$^{+32}_{-23}$ & 3.08$^{+0.07}_{-0.06}$ & 0.74(35) \\
54816 & 0.70$^a$ & 6.74$^a$ & 0.36$^a$ & 5.9$\pm$0.2 & 6.9$^{+0.3}_{-0.2}$ & 1.43$\pm$0.03 & 4.2$\pm$0.2 & 6.61$^{+0.03}_{-0.04}$ & 297$^{+37}_{-36}$ & 4.6$\pm$0.1 & 1.00(34) \\
54817 & 0.70$^a$ & 6.74$^a$ & 0.36$^a$ & 6.1$\pm$0.2 & 6.2$^{+0.4}_{-0.3}$ & 1.51$\pm$0.05 & 3.5$\pm$0.2 & 6.56$^{+0.04}_{-0.05}$ & 318$^{+47}_{-46}$ & 3.3$\pm$0.1 & 0.51(34) \\
54818 & 0.70$^a$ & 6.74$^a$ & 0.36$^a$ & 7.6$^{+0.7}_{-1.0}$ & 10$^{+3}_{-2}$ & 2.01$^{+0.08}_{-0.11}$ & 1.7$^{+0.2}_{-0.3}$ & 6.486$^{+0.002}_{-0.085}$ & 909$^{+175}_{-135}$ & 0.60$^{+0.04}_{-0.02}$ & 0.55(35) \\
54821 & 0.70$^a$ & 6.74$^a$ & 0.36$^a$ & 7.6$^{+0.2}_{-0.3}$ & 4.7$^{+0.6}_{-0.5}$ & 2.20$\pm$0.05 & 4.3$\pm$0.3 & 6.669$^{+0.089}_{-0.001}$ & 502$^{+51}_{-62}$ & 1.00$\pm$0.02 & 0.70(35) \\
54822{\respbf $^b$} & 0.94 & 10.73 & 0.80 & 8.1$\pm$0.2 & 8.7$^{+0.8}_{-0.7}$ & 2.56$^{+0.02}_{-0.03}$ & 13.5$\pm$0.5 & 6.67$\pm$0.01 & 440$^{+26}_{-39}$ & 1.63$\pm$0.01 & 1.17(35) \\
54826 & 0.70$^a$ & 6.74$^a$ & 0.36$^a$ & 6.5$\pm$0.1 & 7.4$\pm$0.2 & 1.63$\pm$0.02 & 7.1$^{+0.3}_{-0.2}$ & 6.63$^{+0.04}_{-0.05}$ & 160$^{+16}_{-14}$ & 5.01$\pm$0.08 & 1.00(35) \\
54827 & 0.70$^a$ & 6.74$^a$ & 0.36$^a$ & 6.7$\pm$0.2 & 6.5$^{+0.4}_{-0.3}$ & 1.79$\pm$0.04 & 5.4$\pm$0.3 & 6.67$\pm$0.06 & 306$^{+23}_{-34}$ & 2.79$^{+0.07}_{-0.06}$ & 0.83(35) \\
54828 & 0.70$^a$ & 6.74$^a$ & 0.36$^a$ & 7.3$\pm$0.2 & 6.9$\pm$0.4 & 1.86$\pm$0.04 & 5.3$\pm$0.3 & 6.63$^{+0.02}_{-0.07}$ & 305$^{+39}_{-27}$ & 2.41$\pm$0.05 & 1.07(35) \\
54829{\respbf $^b$} & 0.94 & 10.73 & 0.80 & 8.5$\pm$0.4 & 11$\pm$2 & 2.62$\pm$0.03 & 18.1$\pm$0.9 & 6.48$^{+0.01}_{-0.08}$ & 485$^{+56}_{-45}$ & 1.98$\pm$0.02 & 1.69(35) \\
54830 & 0.70$^a$ & 6.74$^a$ & 0.36$^a$ & 6.6{\respbf $^c$} & 6.6{\respbf $^c$} & 1.82$\pm$0.08 & 0.7$\pm$0.1 & 6.51$^{+0.06}_{-0.03}$ & 696$\pm$105 & 0.36$\pm$0.01 & 0.87(37) \\
54831 & 0.70$^a$ & 6.74$^a$ & 0.36$^a$ & 8.1$\pm$0.5 & 10$^{+3}_{-2}$ & 2.12$\pm$0.06 & 3.0$^{+0.2}_{-0.3}$ & 6.59$\pm$0.03 & 1061$^{+159}_{-131}$ & 0.85$\pm$0.02 & 0.81(34) \\
54832 & 0.70$^a$ & 6.74$^a$ & 0.36$^a$ & 7.0$\pm$0.3 & 7.1$^{+0.7}_{-0.6}$ & 1.96$\pm$0.05 & 2.9$\pm$0.2 & 6.54$^{+0.04}_{-0.05}$ & 568$^{+56}_{-39}$ & 1.07$\pm$0.02 & 0.77(35) \\
54834 & 0.70$^a$ & 6.74$^a$ & 0.36$^a$ & 6.6$\pm$0.2 & 7.5$\pm$0.2 & 1.55$\pm$0.03 & 4.1$\pm$0.2 & 6.49$\pm$0.04 & 267$^{+15}_{-20}$ & 3.43$\pm$0.07 & 1.07(35) \\
54835 & 0.70$^a$ & 6.74$^a$ & 0.36$^a$ & 6.7$^{+0.1}_{-0.2}$ & 7.2$^{+0.3}_{-0.2}$ & 1.75$\pm$0.03 & 4.5$\pm$0.2 & 6.66$\pm$0.09 & 148$^{+19}_{-16}$ & 2.53$\pm$0.06 & 0.84(35) \\
54836 & 0.70$^a$ & 6.74$^a$ & 0.36$^a$ & 7.8$^{+0.2}_{-0.3}$ & 9.8$^{+0.5}_{-0.4}$ & 1.58$\pm$0.03 & 9.4$^{+0.4}_{-0.3}$ & 6.488$^{+0.086}_{-0.002}$ & 312$^{+36}_{-41}$ & 7.5$\pm$0.2 & 0.79(35) \\
54837 & 0.70$^a$ & 6.74$^a$ & 0.36$^a$ & 6.4$\pm$0.1 & 6.3$\pm$0.1 & 1.73$\pm$0.02 & 5.6$\pm$0.2 & 6.68$^{+0.07}_{-0.01}$ & 236$^{+14}_{-17}$ & 3.26$\pm$0.05 & 0.95(35) \\
54839 & 0.70$^a$ & 6.74$^a$ & 0.36$^a$ & 6.6{\respbf $^c$} & 6.6{\respbf$^c$} & 1.7$\pm$0.2 & 0.41$^{+0.12}_{-0.09}$ & 6.4$\pm$0.1 & 570$^{+202}_{-182}$ & 0.27$\pm$0.02 & 0.94(37) \\
54840 & 0.70$^a$ & 6.74$^a$ & 0.36$^a$ & 6.6$^{+0.6}_{-0.3}$ & 7.8$\pm$0.6 & 1.90$^{+0.03}_{-0.04}$ & 3.8$^{+0.4}_{-0.2}$ & 6.73$^{+0.04}_{-0.13}$ & 362$^{+48}_{-47}$ & 1.59$\pm$0.05 & 0.55(35) \\
54842{\respbf $^b$} & 0.94 & 10.73 & 0.80 & 7.9{\respbf $^d$} & 8.0{\respbf $^d$} & 2.5$\pm$0.1 & 3.3$^{+0.8}_{-0.6}$ & 6.9$\pm$0.2 & 529$^{+217}_{-211}$ & 0.50$\pm$0.03 & 0.55(37) \\
\tableline
\enddata
\tablecomments{The best-fit parameters for the PCA spectra. \\*
$^a$ Out-of-eclipse, the fully covered absorber and partially covered absorber are frozen to the best fit values of the \textsl{Suzaku} out-of-eclipse spectrum (see Table~\ref{Suzaku Spectral Parameters})  \\*
$^b$ Time intervals when 4U 1210-64 is in eclipse.  $N_{\rm H}$ is frozen to the best fit values of the \textsl{Suzaku} eclipse spectrum (see Table~\ref{Suzaku Spectral Parameters}) \\*
$^c$ The Cutoff Energy and Folding Energy parameters at these time intervals are frozen to the weighted average of the out-of-eclipse spectra. \\*
$^d$ The Cutoff Energy and Folding Energy parameters at these time intervals are frozen to the weighted average of the eclipse spectra.}
\label{PCA Blend Spectrum}
\end{deluxetable}

\begin{figure}
\centerline{\includegraphics[width=3.2in]{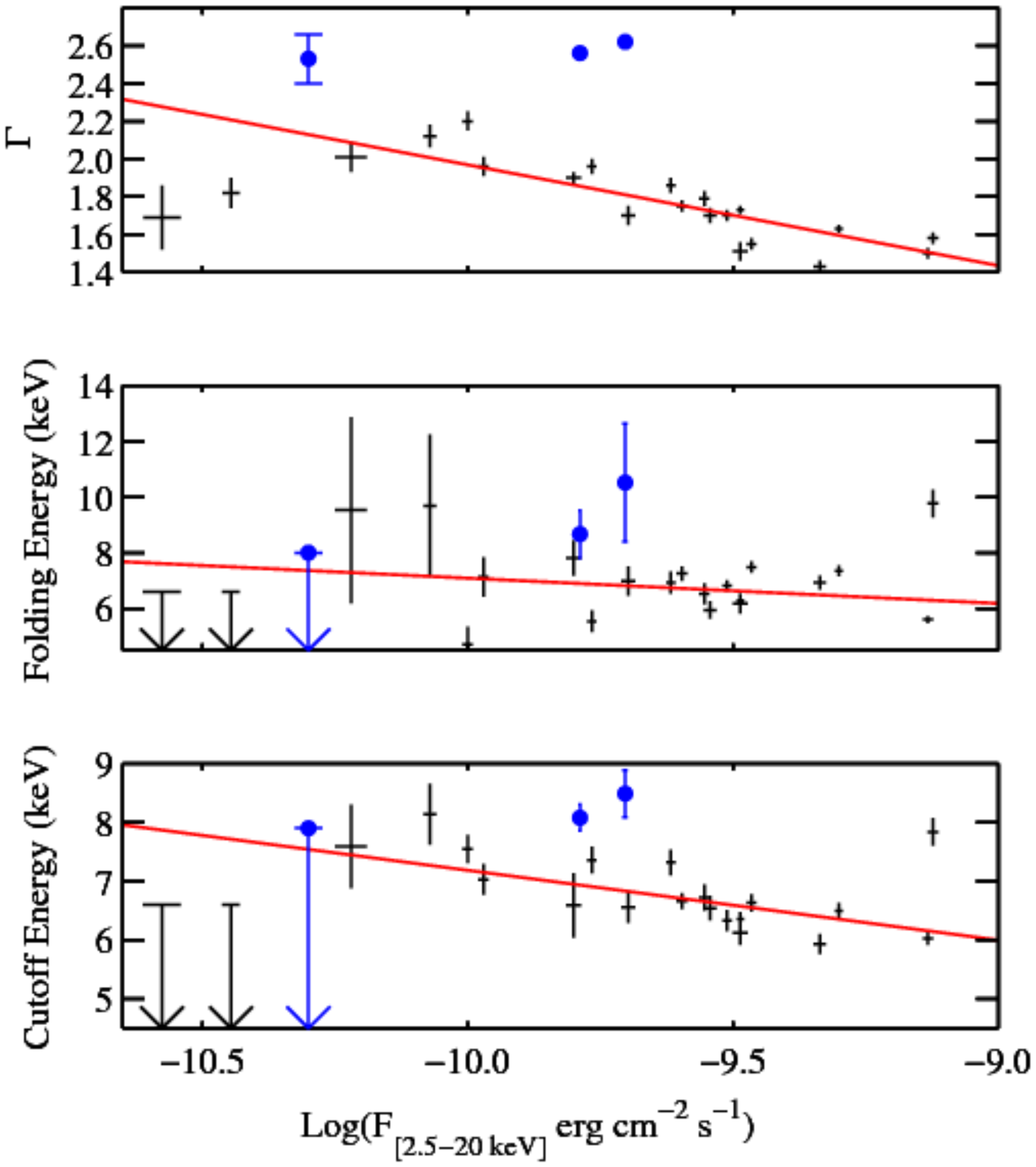}}
\figcaption[june16-2014_PCAspectrum_SpectralParameters.pdf]{
{\mybf Power Law Index (top), Folding Energy (middle) and High Energy Cutoff Energy (bottom) for the PCA spectra vs. the continuum flux in the 2.5--20\,keV band in logarithmic units.  The red lines indicate the best fit for the power law model.  The correlation coefficients (r) are: -0.72 and r=-0.80 for the high energy cutoff and power law index, respectively. The blue points indicate the data collected during eclipse.}
\label{PCA Spectral Parameters}
}
\end{figure}

{\mybf The temporal dependence of the spectral parameters was investigated (see Table~\ref{PCA Blend Spectrum}).  We searched for correlations between the continuum spectral parameters (folding energy, the high energy cutoff and the power law index) with respect to the 2.5--20\,keV continuum flux (see Figure~\ref{PCA Spectral Parameters}).  While no clear correlation between the folding energy and flux was found, the high energy cutoff and power law index are anti-correlated in respect to the continuum flux.  The correlation coefficients (r) are: -0.69 and -0.81 for the high energy cutoff and power law index, respectively.}

\section{Discussion}

In our analysis of 4U 1210-64, we found the presence of an eclipse, long and short-term variability, and an Fe K$\alpha$ emission complex.  Below, we discuss constraints on the mass donor based on the eclipse half-angle, the nature of the compact object, the variability found in the system, and the mechanism responsible for the Fe K$\alpha$ emission seen in 4U 1210-64.

The source emission does not reach 0\,counts s$^{-1}$ in the folded light curves (see Section~\ref{Eclipse Profile}).  Residual emission was also found in the \textsl{Suzaku} and PCA {\respbf observations} (see Tables~\ref{Suzaku Spectral Parameters}--\ref{PCA Blend Spectrum}).  {\respbf While residual emisison has been attributed to a dust-scattering in some other HMXBs \citep[Cen X-3; Vela X-1; OAO 1657-415,][]{1991MNRAS.251...76D,1994ApJ...436L...5W,2006MNRAS.367.1147A}, the residual emission found in 4U 1210-64 is seen at much higher levels compared to the out-of-eclipse emission, and is found in both MAXI (2--20\,keV) and PCA (2.5--20\,keV). A dust-scattering halo is predominantyly a soft X-ray phenomenon: Given the intersteller $N_{\rm H}$ (see Table~\ref{Suzaku Spectral Parameters}) we infer perhaps 10-20$\%$ of the out-of-eclipse flux at 1 keV may be in a dust-scattering halo \citep{1995A&A...293..889P}, but a much smaller fraction in the MAXI and the PCA band.  Since we detect Fe XXV and Fe XXVI lines in eclipse, indicating the presence of an extended region of ionized gas, we believe it's plausible that Compton scattering and reprocessing can account for the residual flux in eclipse \citep{2006ApJ...651..421W}.}

\subsection{Constraints on the Mass-Donor}
\label{Constraints on the Mass-Donor}

Our analysis of the ASM, MAXI and BAT folded light curves reveals the presence of a sharp dip, which is suggestive of an eclipse (see Section~\ref{Eclipse Profile}).

\subsubsection{Eclipse Half-Angle Constraints}

\begin{figure}
\centerline{\includegraphics[width=3.5in]{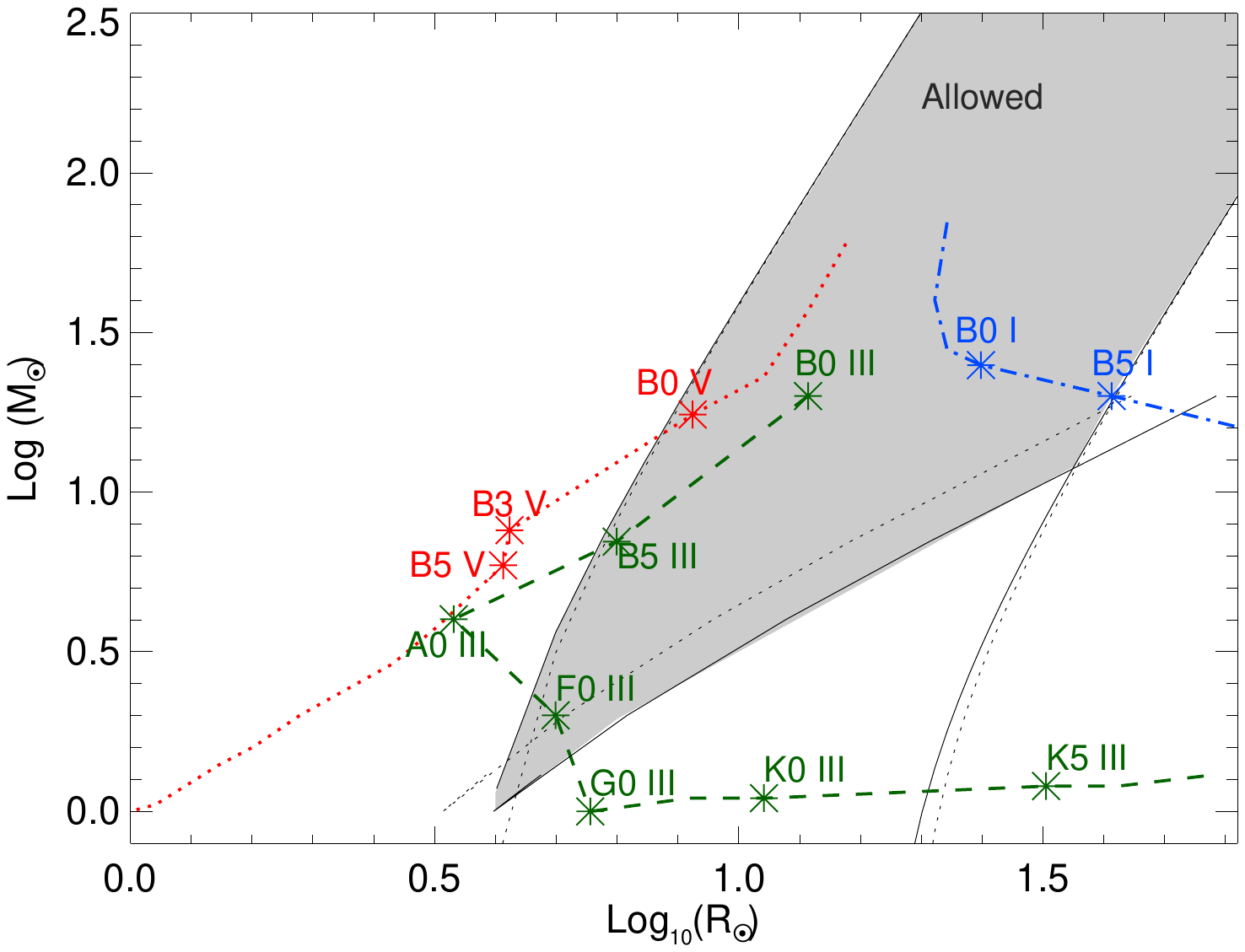}}
\figcaption[march12-2014_mass-constraints.pdf]{
Log-log plot of stellar masses as a function of stellar radii.  The shaded region, derived using Equation E.4 in \citet{1996PhDT........96C} for all possible inclination angles, indicates the allowed spectral types that satisfy the eclipse observed in the BAT, MAXI, and ASM folded light curves provided that the compact object is 1.4 $M_\sun$ (see Figures~\ref{Folded Light Curves} and~\ref{Eclipse Plot}, respectively).  The black dotted lines show the allowed spectral types that satisfy the eclipse for a compact object of 1.9 $M_\sun$ (see Section~\ref{Constraints on the Mass-Donor}).  Stellar masses and radii are given in Table~\ref{Primary Parameters}.
\label{Stellar Constraints}
}
\end{figure}

\begin{figure}
\centerline{\includegraphics[width=3in]{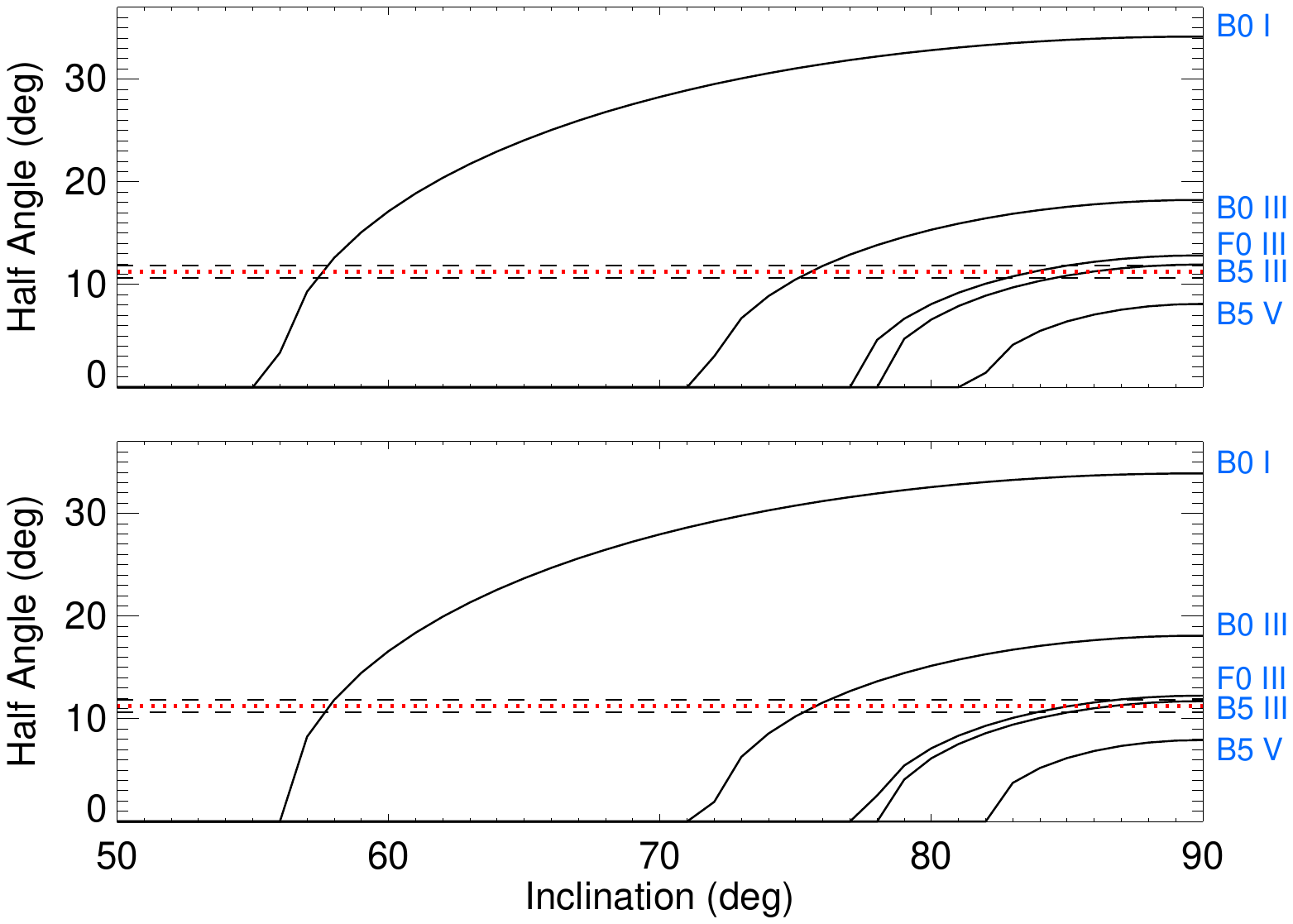}}
\figcaption[4-4-2014_weighted_eclipse_param.pdf]{
The black curves show predicted eclipse half angle as a function of inclination for stars with the indicated spectral types.  The red and black dashed lines indicate the half angle and its estimated error as measured by ASM and MAXI (see Section~\ref{Constraints on the Mass-Donor}).  We assume a neutron star mass of 1.4 $M_\sun$ (top) and of mass 1.9 $M_\sun$ (bottom) and typical masses and radii for the assumed companion spectral type (see Table~\ref{Primary Parameters}).
\label{Half Angle}
}
\end{figure}

X-ray Binaries that are eclipsing have an eclipse duration (see Section~\ref{Eclipse Profile}) that is dependent on the radius of the mass donor, inclination angle of the system and the 
orbital separation of the components.  Using the observed orbital period and Kepler's third law,
the phenomenon can be written in terms of the sum of the donor star and compact object masses,
which stipulates that the eclipse half-angle, $\Theta_{\rm e}$, can now be expressed in terms of
the radius, inclination and masses of the components.  {\respbf In one set of calculations, we
assume a 1.4\,$M_\sun$ compact object which may be appropriate for an accreting neutron star.} 
The region allowed by the measured eclipse half-angle for 4U 1210-64 in the Mass-Radius plot is 
shown in Figure~\ref{Stellar Constraints} (shaded region).  Its inclination is constrained between 
edge-on orbits (left boundary of the shaded region in Figure~\ref{Stellar Constraints}) and close 
to face-on orbits (the right boundary of the shaded region in Figure~\ref{Stellar Constraints}).  
We can attach additional constraints assuming that the mass donor underfills the Roche lobe 
radius, which is dependent on the mass ratio of the system and the orbital separation (see Figure~
\ref{Stellar Constraints}, bottom boundary).  To calculate the eclipse half-angle and the Roche-
lobe radius, we used Equation E.4 in \citet{1996PhDT........96C} and Equation 2 in \citet
{1983ApJ...268..368E}, respectively.  We additionally calculated the minimum inclination angle of 
the system, $i_{\rm min}$, that is consistent with the measured eclipse half-angle (see Table~\ref
{Primary Parameters} and Figure~\ref{Half Angle}).  An intriguing result is that the duration of 
the observed eclipse is inconsistent with the proposed B5 V spectral type (see Figure~\ref{Half 
Angle}).  A B5 V has a mass of 5.9\,$M_\sun$ according to \citet{2006ima..book.....C} and \citet
{2000asqu.book.....A}.  For a donor star of 5.9\,$M_\sun$ to satisfy the eclipse half-angle 
constraint, we calculated the radius must exceed 5.37\,$R_\sun$.  This is clearly larger than the 
radii reported in \citet{2006ima..book.....C} and \citet{2000asqu.book.....A}, which is 4.1\,$R_
\sun$ and 3.9\,$R_\sun$, respectively (see Table~\ref{Primary Parameters}).  Therefore, the radius 
of a B5 V is too small to satisfy our observed eclipse duration (see Figure~\ref{Stellar 
Constraints}).

The eclipse half-angle was calculated as a function of inclination angle for other B-type stars--B0 V, B5 III, B0 III, B5 I and B0 I (see Figure~\ref{Half Angle}).  The eclipse half-angle was found to be consistent with a main-sequence star of spectral class B0 V only at high inclination angles (see Table~\ref{Primary Parameters}).  We also consider intermediate and late spectral types in our analysis (see Figures~\ref{Stellar Constraints} and~\ref{Half Angle}).  These will be discussed in Section~\ref{Intermediate-Mass}.

{\mybf We also calculated the eclipse half-angle as a function of the inclination angle for these stars under the assumption of more massive compact objects.  The results are presented in Figures~\ref{Stellar Constraints} and~\ref{Half Angle} for the scenario of a 1.9\,$M_\sun$ neutron star, which is one of the highest known masses for neutron stars in XRBs \citep{2012ARNPS..62..485L}.  Our results remained the same for more substantial mass donors.  However, the results require slightly higher inclination angles for intermediate and late spectral types (see Figure~\ref{Stellar Constraints}).}

The spectral type of the mass donor places an additional constraint on the distance of 4U 1210-64.  Under the assumption that the R-band magnitude ($m_{\rm R}$) and extinction in the V-band ($A_{\rm V}$) are magnitudes 13.9 and {\mybf 3.3} for a B5 V classification \citep{2009A&A...495..121M}, the distance and average X-ray luminosity of the source are found to be $\sim$2.8\,kpc and 1.79$\pm$0.02 $\times$10$^{35}$\,erg s$^{-1}$, respectively (see Table~\ref{Primary Parameters}).  {\respbf We calculated the extinction in the V-band ($A_{\rm V}$) using Equation 1 in \citet{2009MNRAS.400.2050G} and the measured neutral hydrogen column densities for the fully covered absorber (see Table~\ref{Time Dependent Cutoff Spectrum}).  $A_{\rm V}$ was found to be 3.2$\pm$0.1 for the power law with high energy cutoff model.}

The distance and average X-ray luminosity of 4U 1210-64 assuming the aforementioned spectral types is reported in Table~\ref{Primary Parameters} using the values for $M_{\rm V}$ and $A_{\rm R}$ obtained from \citet{2006ima..book.....C}.  A B0 V star places the system at an estimated distance of {\mybf $\sim$9.5\,kpc} away from the Sun, indicating that 4U 1210-64 could be located in the Carina arm (approximately 10\,kpc).  A supergiant classification places 4U 1210-64 at a galactocentric distance exceeding $\sim$26\,kpc, which is outside the Galaxy.  Therefore, the possibility of a supergiant must be excluded.  {\mybf Since \citet{2009A&A...495..121M}'s previous classification must also be excluded due to the observed eclipse duration}, it is possible that the mass donor could be an early B-type giant or an early F-type giant (see Section~\ref{Intermediate-Mass}).  Main-sequence stars with the exception of very early types and very high inclination angles are also excluded.

\begin{deluxetable}{ccccccccccc}
\rotate
\tablecolumns{11}
\tablewidth{0pc}
\tablecaption{ }
\tablehead{
\colhead{Spectral Type} & \colhead{$M/M_\sun$} & \colhead{$R/R_\sun$} & \colhead{$R_{\rm L}$$/R_\sun$$^a$} & \colhead{$M_{\rm V}$} & {\mybf $(B-R)_{\rm 0}$$^b$} & {\mybf $E(B-R)$$^b$} & \colhead{$i_{\rm min}$$\degr$$^c$} & \colhead{$d_{\rm sun}$$^d$} & \colhead{$d_{\rm gal}$$^e$} & \colhead{$L_{\rm xavg}$} \\
\colhead{} & \colhead{} & \colhead{} & \colhead{} & \colhead{} & \colhead{} & \colhead{} & \colhead{} & \colhead{(kpc)} & \colhead{(kpc)} & \colhead{($\times$10$^{35}$\,erg s$^{-1}$)}} 
\startdata
\textsl{B5 V$^f$} & \textsl{5.9} & \textsl{4.1} & 13.7 & \textsl{-1.2} & {\mybf -0.21} & {\mybf 1.71} & \nodata$^g$ & $\sim$2.8 & $\sim$7.4& $\sim$1.6 \\
\textsl{B5 V$^f$} & \textbf{\textsl{5.9}} & \textbf{\textsl{3.9}} & 13.7 & \textsl{-1.2} & {\mybf -0.21} & {\mybf 1.71} & \nodata$^g$ & $\sim$2.8 & $\sim$7.4& $\sim$1.6 \\
\textsl{B0 V} & \textsl{18} & \textsl{8.4} & 23.1 & \textsl{-4.0} & {\mybf -0.40} & {\mybf 1.90} & 84 & $\sim$9.8 & $\sim$9.3& $\sim$20.0 \\
\textsl{B5 III} & \textsl{7.0} & \textsl{6.3} & 14.9 & \textsl{-2.2} & {\mybf -0.21} & {\mybf 1.71} & 85 & $\sim$4.1 & $\sim$7.3& $\sim$3.4 \\
\textsl{B0 III} & \textsl{20} & \textsl{13} & 24.6 & \textsl{-5.1} & {\mybf -0.34} & {\mybf 1.84} & 75 & $\sim$15 & $\sim$13& $\sim$44.6 \\
\textsl{B5 I} & \textsl{20} & \textsl{41} & 24.6 & \textsl{-6.2} & {\mybf -0.10} & {\mybf 1.60} & 9 & $\sim$29 & $\sim$26& $\sim$175 \\
\textsl{B0 I} & \textsl{25} & \textsl{25} & 27.2 & \textsl{-6.4} & {\mybf -0.36} & {\mybf 1.86} & 57 & $\sim$31 & $\sim$28 & $\sim$197 \\
\textsl{F0 III} & \textsl{2.0} & \textsl{5.0} & 7.7 & \textsl{1.5} & {\mybf 0.45} & {\mybf 1.05} & 82 & $\sim$0.9 & $\sim$7.9 & $\sim$0.2 \\
\textsl{G0 III} & \textsl{1.0} & \textsl{5.7}$^h$ & 5.2$^h$ & \textsl{1.0} & {\mybf 1.11} & {\mybf 0.39} & 77 & $\sim$1.3 & $\sim$7.8 & $\sim$0.3 \\
\enddata
\tablecomments{The values in italics are obtained from \citet{2006ima..book.....C} and \citet{2009A&A...495..121M}. \\*
The values in both italics and bold are obtained from \citet{2000asqu.book.....A} in comparison with those in \citet{2006ima..book.....C}. \\*
$^a$ The definition for the Roche lobe, $R_{\rm L}$, as given in \citet{1983ApJ...268..368E}, assuming $M_{\rm NS}$ is 1.4 $M_\sun$. \\*
$^b$ The value for $(B-R)_{\rm 0}$ was calculated using $(B-V)_{\rm 0}$ and $(R-V)_{\rm 0}$ published in \citet{1994MNRAS.270..229W}. $E(B-R)$ is found by subtracting $(B-R)_{\rm 0}$ from the observed $B-R$ (see Section~\ref{Does the Proposed Mass Donor Spectral Type Agree with Masetti's Optical Spectrum?}) \\*
$^c$ The minimum inclination angle of the system that is consistent with the measured eclipse half-angle. \\*
$^d$ The distance the object is from the Sun. \\*
$^e$ Galactocentric distance of 4U 1210-64. \\*
$^f$ \citet{2009A&A...495..121M}'s proposed spectral type and distance of the object. \\*
$^g$ A B5 V classification is inconsistent with the observed eclipse half-angle. \\*
$^i$ A G0 III classification significantly overfills the Roche-lobe assuming a 1.4 $M_\sun$ compact object.}
\label{Primary Parameters}
\end{deluxetable}

\subsubsection{Does the Proposed Mass Donor Spectral Type Agree with Masetti's Optical Spectrum?}
\label{Does the Proposed Mass Donor Spectral Type Agree with Masetti's Optical Spectrum?}

{\mybf The duration of the eclipse along with the constraint that the mass donor must underfill the Roche lobe allows for the possibility of several different spectral types (see Sections~\ref{Constraints on the Mass-Donor} and~\ref{Intermediate-Mass}).  We compared the expected optical spectra for each proposed mass donor with \citet{2009A&A...495..121M}'s optical spectrum (the top-right panel in Figure 4) to place an additional constraint on the nature of the mass donor.  Since the 4000-5000${\rm \AA}$ region is compressed in \citet{2009A&A...495..121M}'s broadband spectrum, there are possible caveats in the identification of spectral features to correctly classify the mass donor.}

{\mybf The features observed in \citet{2009A&A...495..121M}'s optical spectrum include absorbed Balmer series lines and the emission of neutral helium, singly ionized helium and a blend of doubly ionized nitrogen and carbon.  The optical spectra of B-type stars are expected to show the absorption of neutral helium and H$\alpha$ lines \citep[][and references therein]{2006ima..book.....C}.  Singly ionized calcium features at $\sim$3900${\rm \AA}$ become dominant in F-type stars \citep[][and references therein]{2006ima..book.....C}, which would be difficult to detect in \citet{2009A&A...495..121M}'s broadband spectrum.}  {\respbf The stellar luminosity type can also in principle be determined using spectral lines, which could lead to distinguishing a B III or B V type.  For B5 stars, the ratio of Si II to Si III as well as the Al III and Fe III lines can be used to determine luminosity type \citep{2009ssc..book.....G}.  Unfortunately, the existing optical spectra are not suitable in detecting these effects.}

The observed value of the $B-R$ color of 1.5\footnote{http://www.iasfbo.inaf.it/\~masetti/IGR/main.html} was compared with the intrinsic $(B-R)_{\rm 0}$ for the proposed mass donors (see Table~\ref{Primary Parameters}).  Calculating the difference between the observed $B-R$ and the intrinsic $(B-R)_{\rm 0}$, we found the reddening values $E(B-R)$ for each proposed spectral type for the mass donor (see Table~\ref{Primary Parameters}).  We calculated the reddening in the $B-V$ band, $E(B-V)$, using Equation 1 in \citet{2009MNRAS.400.2050G} and the measured neutral hydrogen column densities for fully covered absorber (see Table~\ref{Time Dependent Cutoff Spectrum}).  We converted $E(B-V)$ to $E(B-R)$ using Table 3 in \citet{1985ApJ...288..618R}.  We found $E(B-R)$ to be 1.82$\pm$0.08 for the power law with high energy cutoff model, where the reddening was found to be consistent with main-sequence and giant B-type stars but was inconsistent with F-types.  Since a B5 V classification does not satisfy the eclipse half-angle, the optical features might indicate that the mass donor is a B-type giant.  We note the optical spectra of B-type giants would be difficult to distinguish from that of main-sequence B-type stars because of the low-resolution of \citet{2009A&A...495..121M}'s broadband spectrum.  F-type giants cannot be completely excluded due to systematic effects which prevent a definite determination.  For example, we assumed that the fully covered $N_{\rm H}$ is entirely interstellar in origin (see Section~\ref{Suzaku Spectral Analysis}).  {\newbf While this is the simplest interpretation of the data, we cannot exclude the possibility that the fully covered $N_{\rm H}$ is due to a combination of intrinsic and interstellar absorbers. According to our calculation, if $\sim$40$\%$ of the measured fully covered $N_{\rm H}$ is intrisic to the source and the rest interstellar, then the inferred $E(B-R)$ would be consistent with an F-type mass donor.}

{\mybf Finally, we considered the possibility that the apparent spectral type is affected by heating by the radiation of the X-ray source.  We first calculated the flux of 4U 1210-64 using both \textsl{Suzaku} and the PCA.  Out-of-eclipse, the flux of 4U 1210-64 was found to be $\sim$2$\times$10$^{-10}$\,erg cm$^{-2}$ s$^{-1}$ for the power law model with a high energy cutoff model in the \textsl{Suzaku} data.  The flux in the PCA is reported in Table~\ref{PCA Blend Spectrum} ranging from 0.29$\pm$0.03$\times$10$^{-10}$\,erg cm$^{-2}$ s$^{-1}$ to 8.9$\pm$0.4$\times$10$^{-10}$\,erg cm$^{-2}$ s$^{-1}$.  We converted the apparent magnitude in the R band \citep{2009A&A...495..121M} into an optical flux using \citet{1979PASP...91..589B}.  The optical flux in the R band was found to be 0.08\,Jy, which can be converted to $\sim$2$\times$10$^{-7}$\,erg cm$^{-2}$ s$^{-1}$.  The flux ratio $F_{X}/F_{opt}$ was found to be $\sim$10$^{-4}$, which is much smaller than what is observed in systems were irradiation is important \citep[e.g. $F_{X}/F_{opt}$ was found to be $\sim$10$^{2}$ in Her X-1;][]{1983ARA&A..21...13B}.  We conclude that irradiation effects are negligible in 4U 1210-64.}

\subsubsection{Is 4U 1210-64 an Intermediate-Mass X-ray Binary?}
\label{Intermediate-Mass}

{\mybf In addition to the possibility that the donor star is a B-type giant, the duration of the eclipse suggests that an Intermediate-Mass classification cannot be ruled out}.  We found the eclipse half angle is consistent with F0 and G0 giants at inclination angles exceeding 79$\degr$ and 70$\degr$, respectively (see Table~\ref{Primary Parameters}).  Using the values for $M_{\rm V}$ and $A_{\rm R}$ published in \citet{2006ima..book.....C} and \citet{2009A&A...495..121M}, the distance of 4U 1210-64 was found to be $\sim$0.9\,kpc and $\sim$1.3\,kpc for a mass donor of spectral type F0 III and G0 III, respectively; which places 4U 1210-64 at a luminosity of $\sim$10$^{34}$\,erg s$^{-1}$ (see Table~\ref{Primary Parameters}).  Other intermediate XRBs that host F-type stars include Cyg X-2 and Her X-1 \citep[e.g.][]{1995exru.book.....S}, which have considerably higher luminosities on the order of $\sim$10$^{37}$\,erg s$^{-1}$.  The luminosities calculated for the F0 III and G0 III spectral types still exceed that of cataclysmic variables (see Section~\ref{What is the Nature of the Compact Object?}).

The possibility that the mass donor in 4U 1210-64 is an intermediate or late-type star hints at the presence of an accretion disk.  XRBs that host intermediate- and low-mass stars accrete matter through Roche-lobe overflow \citep[see Equation 2,][]{1983ApJ...268..368E}.  The Roche-lobe places an additional constraint on the spectral type of the mass donor in 4U 1210-64.  While a spectral type of G0 III satisfies the observed eclipse half-angle, the Roche-lobe would be significantly overfilled (see Table~\ref{Primary Parameters}).

Finally, we discuss caveats in our hypothesis that the mass donor is an Intermediate-Mass star.  No strong disk component was found, which is expected in the spectra of Intermediate-Mass X-ray Binaries \citep[e.g.][]{1995exru.book.....S}.  We note that the reddening,  $E(B-R)$, was found to be consistent with main-sequence and giant B-type stars but was inconsistent with {\respbf F-type and G-type stars} (see Section~\ref{Does the Proposed Mass Donor Spectral Type Agree with Masetti's Optical Spectrum?}).  While $E(B-R)$ is apparently inconsistent with F-type stars, the possibility of systematic effects prevents excluding F-type stars as the possible mass donor {\newbf (see Section~\ref{Does the Proposed Mass Donor Spectral Type Agree with Masetti's Optical Spectrum?})}.  However, we can confidently exclude G-type stars due to the additional constraint that the Roche-lobe is not significantly overfilled.

\subsection{What is the Nature of the Compact Object?}
\label{What is the Nature of the Compact Object?}

The nature of the compact object present in 4U 1210-64 remains ambiguous.  An analysis of the ASM, MAXI and PCA power spectra shows that no pulsation period could be identified.  {\mybf The PCA power spectra, which cover the range of 860\,s--38\,d (see Figure~\ref{Low Frequency}) and 10\,ms--14\,minutes (see Figure~\ref{Short Power}), are dominated by red noise, which could compromise our search for the pulsation period.  While we removed the low-frequency noise from the power spectra (see Section~\ref{PCA Temporal}), the PCA power spectrum covering the range of 860\,s--38\,d was still compromised due to the orbital period of \textsl{RXTE} (see Figure~\ref{Low Frequency})}.  A pulsation period would provide a clear indication that the compact object is a neutron star.  Spectral results so far have also been inconclusive.  Cyclotron lines, which would have proved a neutron star explanation for the compact object, are absent in the \textsl{Suzaku} spectra.  Additionally, {\respbf the \textsl{Suzaku} and PCA data suggest the continuum can be modeled using an absorbed cutoff power law where the high-energy cutoff is 5.5$\pm$0.2\,keV}. Since a firm identification of the nature of the compact object in 4U 1210-64 has so far proven elusive, we compare our findings to systems where the compact object is known.

We first discuss the possible scenario that the compact object present in 4U 1210-64 is a black hole.  Observations show that the exponential cut-off energy in HMXBs that host black holes exceeds 60\,keV \citep{2009ApJ...694..344T}, sharply contrasting with the 5.5$\pm$0.2\,keV cut-off observed in the \textsl{Suzaku} spectra.  Additionally, \textsl{INTEGRAL} observations reveal the presence of soft excess in 4U 1210-64 \citep{2010ApJ...511..A48}, which is characteristic of HMXBs that host neutron stars \citep{2004ApJ...614..881H}.  The low high-energy cutoff suggests that a black hole explanation of the compact object is unlikely.

We also consider the possibility that the compact object could be an accreting white dwarf.  The luminosities observed in cataclysmic variables depend on the magnetic nature of the white dwarf, which affects the mode of accretion.  The most luminous sub-type of CV, the intermediate polars, were found to be on the order of 10$^{31}$--4$\times$10$^{33}$\,erg s$^{-1}$ \citep{2009A&A...496..121B}.  In comparison, the luminosities calculated for 4U 1210-64 exceed the above result by at least 1--2 orders of magnitude (see Table~\ref{Primary Parameters}).  Assuming a bremsstrahlung fit, \citet{2009A&A...496..121B} found the temperatures ($kT_{\rm brems}$) of intermediate polars are on the order of 10--40\,keV, which differs from the value observed in 4U 1210-64 {\newbf (see Section~\ref{Suzaku Spectral Analysis})}.  Since a bremsstrahlung model was found to be an unsatisfactory fit to the PCA data, we conclude that the CV explanation is unlikely.

Finally, we discuss the possibility that 4U 1210-64 contains a neutron star.  Several geometries have been proposed to describe the apparent lack of a signal corresponding to the pulsation period.  One possibility is a co-alignment of the magnetic and spin axes of the neutron star \citep[e.g.][]{2010ApJ...719..451B}.  A second explanation suggests that throughout the rotation the accretion beam points in our direction \citep[e.g.][]{2010ApJ...719..451B}.  The absence of a well defined pulsation period could also be explained by a weak magnetic field.  Another possibility is that the compact object in 4U 1210-64 is a slowly rotating neutron star \citep[e.g. 2S 0114+650,][and references therein]{2008MNRAS.389..608F}.

\subsection{What is the physical process responsible for the low state observed in the ASM data?}
\label{What is the physical process responsible for the low state observed in the ASM data?}

The ASM data reveal the presence of three distinct system states as previously noted by \citet{2008ATel.1861....1C}.  These are two active states and one low state we interpret as quiescence (see Section~\ref{ASM Temporal Analysis}).  This long term variability is suggestive of a variable accretion rate.  

We first discuss the possibility that 4U 1210-64 is powered by the Be mechanism.  {\mybf In BeXBs the compact object accretes material from the circumstellar decretion disc of a rapidly rotating main-sequence or giant B-type star.}  If the system is a BeXB, changes in the circumstellar disc around the Be star could also explain the period of low activity.  Observations indicate that bright and faint states might correspond to the formation and dissipation of the circumstellar disc \citep{2010MNRAS.401...55R}.  The time-scale for the development and disappearance of a circumstellar disc is typically on the order of 3--7\,years, which is consistent with the ASM data.  One notable BeXB is SAX J2103.5+4545, which consists of a neutron star in a 12.7\,day orbit around a B0Ve star \citep{2010MNRAS.401...55R}. The timescale for the development and disappearance of the circumstellar disc is 1--2\,years, possibly due to the short orbital period \citep{2010MNRAS.401...55R}.  {\mybf While a BeXB explanation supports the presence of high and low states observed in the ASM data, the presence of an eclipse of the compact object is inconsistent with most main-sequence B-type stars, where main-sequence stars later than a B0 do not satisfy the observed eclipse half angle (see Figures~\ref{Stellar Constraints}--~\ref{Half Angle}).  This is not surprising due to the smaller radius of the mass donors observed in most BeXBs.  We note; however, that B-type giants would satisfy the observed eclipse duration at high inclination angles (see Section~\ref{Constraints on the Mass-Donor}).  Additionally, the Balmer lines, particularly the H$\alpha$ line, were found to be in absorption during \citet{2009A&A...495..121M}'s optical campaign of 4U 1210-64 (MJD\,54529.3).  Since Masetti's observation occured during a high state of the system (see Table~\ref{ASM State Table} and Section~\ref{ASM Temporal Analysis}), we would expect to see H$\alpha$ in emission \citep{2010ApJS..187..228S}.}

{\mybf We also consider the possibility that the state transitions originate due to a mechanism similar to what is observed in Black Hole Candidates (BHCs).  Multiple states are observed in BHCs, which are defined as soft/high, intermediate and hard/low \citep[Cygnus X-1; GX 339-4,][]{2013A&A...554A..88G,2006smqw.confE...1N,2010MNRAS.403...61D}.  The high-energy cutoff in BHCs exceeds 60\,keV \citep{2009ApJ...694..344T}, which is in variance with our results obtained with \textsl{Suzaku} and PCA.  Additionally, a disk blackbody is required to model the low energy spectra of BHCs in the soft/high state \citep{1973A&A....24..337S,2006smqw.confE...1N}, which is not seen in the \textsl{Suzaku} and PCA analysis of 4U 1210-64.} 

Finally, we discuss the possibility that the mode of accretion in 4U 1210-64 is Roche-lobe overflow, which primarily occurs in both Intermediate and Low-Mass X-ray Binaries.  {\mybf We compare the long-term behavior of 4U 1210-64 to that seen in soft X-ray transients, NS-LMXBs that host at least two different states \citep[e.g. Aql X-1;][]{2013arXiv1308.6091S}}.  In soft X-ray transients, the physical mechanism that could lead to a reduction of intensity is changes in the accretion rate, $\dot{M}$ \citep{2013arXiv1308.6091S}.  The luminosity is significantly reduced when the accretion rate is low.  Since the magnetic field of the majority of NS-LMXBs is weak, the propeller effect becomes important when the accretion rate, $\dot{M}$, is low \citep{2013arXiv1308.6091S}.  The long-term behavior of 4U 1210-64 differs from soft X-ray transients.  While there are extended low states \citep{2013PASJ...65...26M,2013arXiv1308.6091S}, the high states are shorter and brighter than what is observed in 4U 1210-64 {\mybf \citep{2014MNRAS.438.2634C}}.

{\mybf No mechanisms described above appear to be consistent with 4U 1210-64.  While there are uncertainties in the spectral classification, we note that the behavior of Intermediate-Mass X-ray Binaries is not well known.  Therefore, if 4U 1210-64 is a member of this class, unusual variability may be possible.}

\subsection{What is the origin of the variability in the high-state?}
\label{What is the origin of the variability in the high-state?}

Our analysis of the \textsl{Suzaku} {\mybf and PCA} data reveals the presence of strong variability in the light curves ({\mybf see Figures~\ref{Suzaku Light Curve} and~\ref{PCA Light Curve})}.  {\mybf In the \textsl{Suzaku} data, ``flares" were observed to reach nearly 1.4 times the mean count rate (i.e. a modulation depth of 140$\%$)}.  {\mybf The variability is even stronger in the PCA data, where the modulation depth was found to be 330$\%$.  A reduced count rate was found in the PCA light curve between $\sim$MJD\,54830--54833, which is outside the eclipse (see Figure~\ref{PCA Light Curve}).}  The unabsorbed flux of 4U 1210-64 was found to vary by a factor of $\sim$25 over the course of both the \textsl{Suzaku} and {\mybf PCA observations (see Sections~\ref{Suzaku Spectral Analysis} and~\ref{PCA Spectral Description})}.

\begin{figure}
\centerline{\includegraphics[width=3.5in]{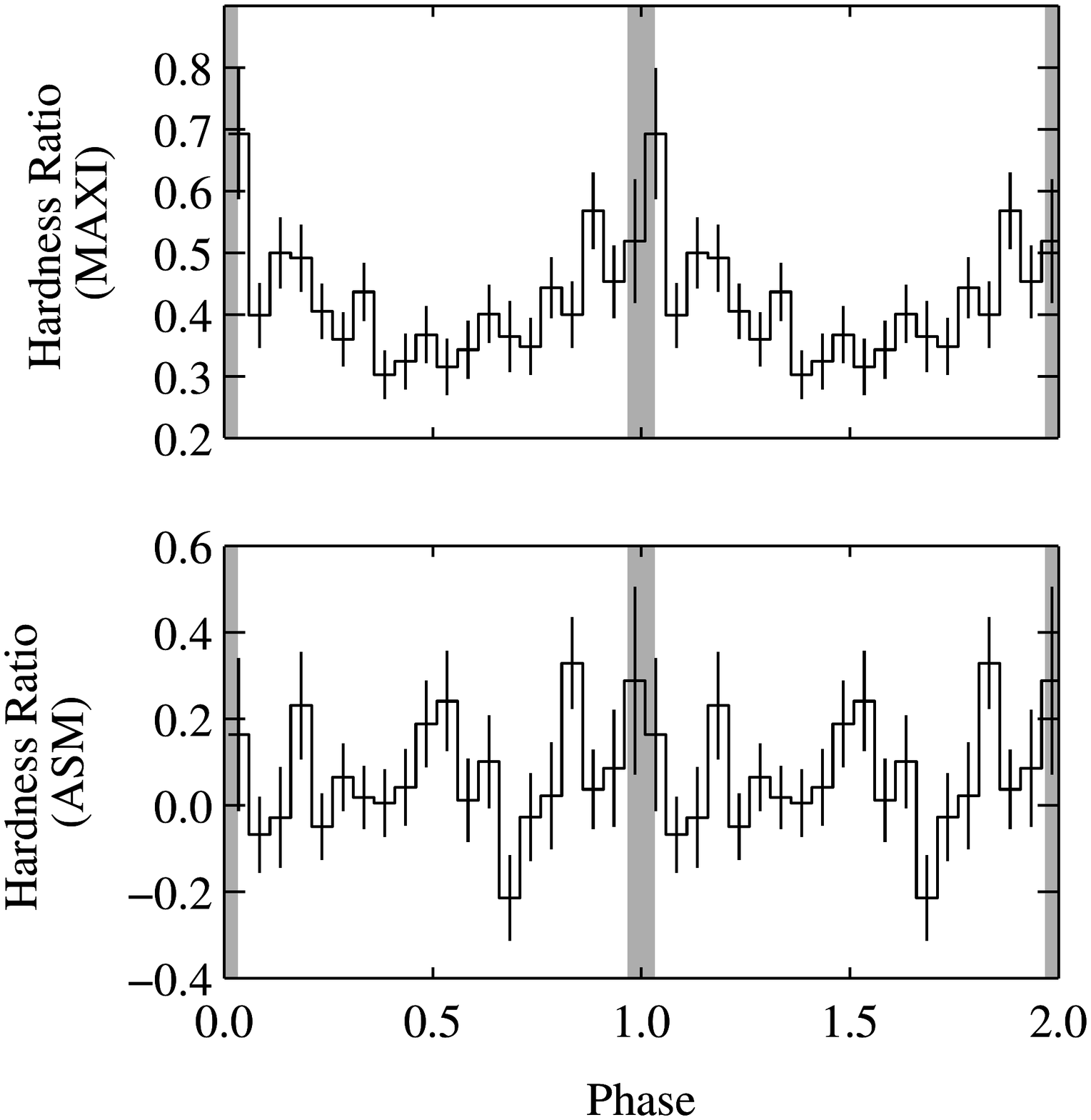}}
\figcaption[november06-2013_hardness-ratios.pdf]{
Hardness ratios of the folded MAXI (top) and ASM (bottom) light curves.  The hardness ratio is defined as $C_{\rm hard}$-$C_{\rm soft}$/$C_{\rm hard}$+$C_{\rm soft}$, where the soft and hard energy bands are defined as 2--4\,keV and 4--10\,keV and 1.5--5\,keV and 5-12\,keV for the MAXI and ASM light curves, respectively.  The eclipse is indicated by the shaded regions.
\label{Folded Hardness Ratios}
}
\end{figure}

{\mybf The large variability in the Suzaku and PCA light curves could be attributed to several different physical processes, all resulting in changes in the accretion rate, $\dot{M}$. A positive correlation between the \textsl{Suzaku} hardness ratio and continuum flux was found (see Figure~\ref{Hardness Correlation}), which provides evidence against a strong wind.  This is further strengthened by a decrease in the \textsl{Suzaku} hardness ratio during the egress phase of the observation.  Since the system is eclipsing, a strong wind should lead to an increase in absorption and thus the X-ray hardness ratio during the ingress and egress phases of the orbit \citep{1988ApJ...324..974C,2013A&A...554A..37D,2008ApJ...675.1487S}.  The folded MAXI and ASM light curves (see Figure~\ref{Folded Hardness Ratios}, top and bottom) provide additional evidence against it.  The increase in the folded MAXI hardness ratio is modest (Figure~\ref{Folded Hardness Ratios}, top), indicating the possible presence of a tenuous wind but not of a typical HMXB wind \citep[e.g. Vela X-1; Cen X-3,][]{2013A&A...554A..37D,2008ApJ...675.1487S}.  Such behavior is not seen in the folded hardness ratio produced by the ASM, which is possibly due to the low count rate (see Figure~\ref{Folded Hardness Ratios}, bottom).  The observed $N_{\rm H}$ could possibly originate in an accretion stream \citep[e.g. Cygnus X-1,][and references therein]{2009ApJ...690..330H}.}

Magnetic and centrifugal barriers have been proposed to inhibit the accretion process in XRBs on hourly timescales \citep{2008arXiv0811.0995B}, where the timescale is consistent with the variability in the light curve.  {\mybf This could explain the reduced count rate in the PCA lightcurve (see Figure~\ref{PCA Light Curve}, top).}  Another mechanism that could lead to {\mybf such a reduction} is the formation and dissipation of an unstable accretion disk \citep[e.g.][and references therein]{2011ApJ...727...59B}.

\subsection{Emission Lines}
The analysis of spectral data produced by \textsl{Suzaku} reveals the presence of emission lines at energies 2.6\,keV, 6.4\,keV, 6.7\,keV and 6.97\,keV.  We interpret the emission lines as S XVI K$\alpha$, Fe K$\alpha$, Fe XXV K$\alpha$ and Fe XXVI K$\alpha$, respectively.  Below we discuss the mechanisms that are proposed to explain the emission lines seen in 4U 1210-64.

\subsubsection{Fe K$\alpha$ Emission}
A 6.4\,keV emission line has been shown to be present in many XRBs \citep[e.g. Vela X-1, 4U 1700-377, Cen X-3 and 4U 1822-37;][] {2002ApJ...564L..21S,2005A&A...432..999V,2011ApJ...737...79N,2013arXiv1311.4618S}.   The origin of the 6.4\,keV emission line is due to neutral Fe or Fe in a low ionization state.  Unless otherwise stated, we assume that Fe I emission is responsible for the 6.4\,keV emission line.

Our analysis of the Fe I line indicates that the line flux tracks the flux in the continuum in the 7.1--9.0\,keV band when 4U 1210-64 is out-of-eclipse and is not detected in eclipse (see Section ~\ref{Suzaku Spectral Analysis}).  {\mybf This is suggestive that the region responsible for the Fe I emission is close to the compact object.  The accretion mechanism in 4U 1210-64 differs from the X-ray excited wind observed in the HMXB Cen X-3 \citep{1993ApJ...403..322D}, but the Fe K$\alpha$ emission region is similar in both objects.  For instance, the Fe I emission line in Cen X-3 was observed to be weakest during the eclipse phase \citep{1996PASJ...48..425E,2012BASI...40..503N}.  Because of these similarities, we compare 4U 1210-64 with Cen X-3 to understand the Fe K$\alpha$ emission feature}.  

Possible mechanisms that have been suggested to cause the fluorescence of cold material include a plasma layer at the surface of the Alfven shell and an optically thick accretion disc \citep{1980A&A....87..330B,1996PASJ...48..425E}.  The flux of Fe I was found to decrease by more than an order of magnitude during eclipse, which is a comparable to the change of flux observed in the continuum.  This is further indication that the origin of the Fe I emission is close to the compact object.  The slope of the Fe I flux versus the continuum flux is near unity, which shows that the Fe I emission is in agreement with fluorescence \citep{2012A&A...547A.103N,2013A&A...551A...1R}.

We also consider the mechanism responsible for the Fe XXV and Fe XXVI emission features.  A possible correlation between the flux of both the Fe XXV and Fe XXVI emission features with respect to the continuum flux in the 7.1--9.0\,keV band was found, which shows that the presence of Fe XXV and Fe XXVI increases as the continuum flux increases (see Section 3.4).  We note that the slope between the continuum flux and the flux of the lines is significantly less than 1, which we interpret as a possible sign that an increasing part of the medium might be completely ionized.

To place constraints on the state of the plasma, we analyzed the flux ratio between the Fe XXV and Fe XXVI emission features and the continuum flux in the 7.1--9.0\,keV band.  No change was found in the flux ratio between the Fe XXV and Fe XXVI emission features and the continuum flux in the 7.1--9.0\,keV band (see Table~\ref{Time Dependent Cutoff Spectrum}), which indicates that the {\respbf Fe XXV and Fe XXVI features possibly originate in the same region of a structured medium, where the} ionization state can be independent of luminosity \citep{1996PASJ...48..425E}, which is in agreement with our result of a constant flux ratio as a function of luminosity.

Fe XXVI and Fe XXVI are likely due to photoionization.  Recombination followed by electron cascade transitions is present in systems such as SMC X-1 \citep{2001ApJ...563L.139V}, Cen X-3 \citep{2012A&A...547A.103N,2013A&A...551A...1R}, Vela X-1 \citep{2004AJ....127.2310G}, 4U 1700-37 \citep{1996ApJ...468L..33L} and 4U 1822-37 \citep{2013arXiv1311.4618S}.  Emission features are more prominent during eclipse since direct emission from the compact object irradiating the accretion stream is no longer visible.  The photoionization mechanism must originate in regions of low density since the range of luminosities we inferred in 4U 1210-64 (see Table~\ref{Primary Parameters}) is significantly lower than what is observed in systems such as Cen X-3 and SMC X-1 \citep[$\sim$10$^{37}$\,erg s$^{-1}$,][]{2012BASI...40..503N,2001ApJ...563L.139V}.

The very large EQWs of the Fe XXV and Fe XXVI emission features in eclipse (see Section 3.4) are consistent with an origin due to the reprocessing of photons in the accretion stream.   In comparison, the EQW of the Fe XXV and Fe XXVI emission features observed in the HMXB Cen X-3 is largest during eclipse and tends to decrease as the continuum flux increases \citep{2012BASI...40..503N}.  \citet{2012BASI...40..503N} show that the region responsible for the Fe XXV and Fe XXVI emission observed in Cen X-3 is extended and is comparable to the size of the mass donor in the system, which is likely what we observe in 4U 1210-64.

\subsubsection{The presence of S XVI in 4U 1210-64}
Different ionization species of low to mid-Z elements such as S are present in eclipsing XRBs \citep[e.g. Vela X-1, 4U 1700-377 and LMC X-4,][]{2002ApJ...564L..21S, 2005A&A...432..999V, 2010ApJ...720.1202H}.  While near neutral fluorescent lines in addition to highly ionized species of emission lines were observed in SGXBs such as Vela X-1 \citep{2002ApJ...564L..21S}, the \textsl{Suzaku} spectra of 4U 1210-64 reveal only the highly ionized species of S XVI.

It has been shown that photoionization and radiative recombination are responsible for the presence of hydrogen-like species of S in absorbed XRBs \citep{2004ApJ...600..358I}.  The S XVI K$\alpha$ emission features only appeared during part of the \textsl{Suzaku} observation (see Table~\ref{Time Dependent Cutoff Spectrum}).  As a result, we could not measure the temporal variability of the EQW or fluxes of S XVI.

\section{Conclusion}

{\respbf 4U 1210-64, for which we determined an orbital period of 6.7101$\pm$0.0005\,days, is a unique XRB. The companion star was previously proposed to have a spectral type of B5 V. We found that a B5 V classification does not satisfy the eclipse half-angle, compelling evidence against a B5 V classification. 4U 1210-64’s spectral features seem to indicate that the mass donor could be a B0 V or B0-5 III star. A Be-type accretion mechanism, with most BeXBs hosting primaries of spectral type late O to B2 \citep{1998A&A...338..505N}, is unlikely, since these systems usually have longer periods and are transients. A supergiant classification must be excluded since the implied distance would put the object outside the Galaxy.  F-type giants also satisfy the constraints imposed by the eclipse half-angle and Lagrange point, L1, where Roche-lobe transfer would be expected to occur.  To further constrain the spectral type of the mass donor, the reddening values $E(B-R)$ were calculated for the possible spectral types. The reddening was found to be consistent with main sequence and giant B-type stars but was inconsistent with F-types.  Due to the uncertainties in the conversion between the $N_{\rm H}$ and $E(B-R)$, we cannot completely exclude an F-type mass donor.}

4U 1210-64 hosts a compact object that remains ambiguous in nature.  No signs of pulsations or cyclotron features were found in our analysis of 4U 1210-64, which would prove that the compact object is a neutron star.  The spectral properties of the continuum strongly contrast with those typically seen in black hole candidates (BHC) as well as CVs.  {\respbf In particular, a disk blackbody is required to model the low energy spectra of BHCs in the soft/high state while CVs are typically fit with a bremsstrahlung model.}  While the nature of the compact object has proven elusive, a neutron star with a weak magnetic field possibly aligned with the spin axis is consistent with the lack of pulsations and cyclotron features.

Emission lines at 2.62\,keV, 6.41\,keV, 6.7\,keV and 6.97\,keV were all clearly detected in the \textsl{Suzaku} spectra, which we interpret as S XVI K$\alpha$, Fe K$\alpha$, Fe XXV K$\alpha$ and Fe XXVI K$\alpha$, respectively.  The flux of the Fe K$\alpha$ lines closely tracks the flux of the unabsorbed continuum.  We found a linear relationship between the flux of Fe I vs. the continuum, which shows that the most probable origin of the Fe I line is fluorescence of cold and dense material close to the compact object.  An origin close to the compact object is further supported by the fact that Fe I is not clearly detected during eclipse.  The slopes of the relationship between the logarithm of the Fe XXV and Fe XXVI flux versus the logarithm continuum possibly show that an increasing part of the medium might be completely ionized as the flux increases.

{\respbf Strong variability was found in the \textsl{Suzaku} and PCA observations.  The out-of-eclipse flux was found to be 1.73$^{+0.06}_{-0.05}$$\times$10$^{-10}$\,erg s$^{-1}$ cm$^{-2}$, which implies a luminosity $\sim$10$^{34}$-$\times$10$^{36}$\,erg s$^{-1}$. The variability was found to be a factor of 25 in both the \textsl{Suzaku} and PCA observations. A positive correlation was seen in both the \textsl{Suzaku} hardness-intensity diagram and the PCA color-color diagram, which provides evidence against a strong wind.  In eclipsing X-ray binaries, a strong wind should lead to an increase in absorption and thus the X-ray hardness ratio during the ingress and egress phases of the orbit.  Additional evidence from the folded MAXI and ASM light curves suggests that the observed $N_{\rm H}$ could originate possibly originate in an accretion stream.}

{\respbf 4U 1210-64 appears to be a NS-HMXB but conclusive evidence remains to be found.  Additional multi-wavelength observations are required to achieve a full understanding of the source.}
\acknowledgements

We thank Drs. Tim Kallman, Vanessa McBride and Joern Wilms for useful discussion.  We also thank the anonymous referee for useful comments.

\end{document}